\documentclass[12pt, draftclsnofoot, onecolumn]{IEEEtran}
\ifCLASSINFOpdf
 \usepackage[pdftex]{graphicx}
\else
  \usepackage[dvips]{graphicx}

\fi

\usepackage{graphicx}
\usepackage{epsfig}
\usepackage[cmex10]{amsmath}
\usepackage{amsfonts}
\usepackage{amssymb}
\usepackage{amsthm}
\usepackage{morefloats}
\usepackage{cite}
\usepackage{bm}
\usepackage{hyperref}

\usepackage{algorithm} 
\usepackage{algpseudocode}

\usepackage[caption=false,font=normalsize,labelfont=sf,textfont=sf]{subfig}
\begin{document}
%
\title{Downlink Rate Analysis for Virtual-Cell based Large-Scale Distributed Antenna Systems}
\author{Junyuan~Wang,~\IEEEmembership{Student Member,~IEEE},~and~Lin~Dai,~\IEEEmembership{Senior Member,~IEEE} \thanks{J. Wang and L. Dai are with the Department of Electronic Engineering, City University of Hong Kong, 83 Tat Chee Avenue, Kowloon Tong, Hong Kong, China (email: eewangjy@gmail.com; lindai@cityu.edu.hk). }}
\maketitle

\begin{abstract}
Despite substantial rate gains achieved by coordinated transmission from a massive amount of geographically distributed antennas, the resulting computational cost and channel measurement overhead could be unaffordable for a large-scale distributed antenna system (DAS). A scalable signal processing framework is therefore highly desirable, which, as recently demonstrated in \cite{Dai_TWireless}, could be established based on the concept of virtual cell.

In a virtual-cell based DAS, each user chooses a few closest base-station (BS) antennas to form its virtual cell, that is, its own serving BS antenna set. In this paper, we focus on a downlink DAS with a large number of users and BS antennas uniformly distributed in a certain area, and aim to study the effect of the virtual cell size on the average user rate. Specifically, by assuming that maximum ratio transmission (MRT) is adopted in each user's virtual cell, the achievable ergodic rate of each user is derived as an explicit function of the large-scale fading coefficients from all the users to their virtual cells, and an upper-bound of the average user rate is established, based on which a rule of thumb is developed for determining the optimal virtual cell size to maximize the average user rate. The analysis is further extended to consider multiple users grouped together and jointly served by their virtual cells using zero-forcing beamforming (ZFBF). In contrast to the no-grouping case where a small virtual cell size is preferred, it is shown that by grouping users with overlapped virtual cells, the average user rate can be significantly improved by increasing the virtual cell size, though at the cost of a higher signal processing complexity.

\end{abstract}


\vspace{5mm}

\section{Introduction}

The fast growing demand for data rate has posed unprecedented challenges for designers of next-generation (5G) mobile communication systems. To support a large number of mobile users with high-data-rate applications such as online gaming and streaming high-definition video, many new innovative solutions have been proposed (see \cite{JAndrews1} for a comprehensive overview). In general, there are two main ways to improve the data rate:
\begin{enumerate}
\item Utilizing more bandwidth: By moving to the millimeter wave (mmWave) spectrum which was widely regarded as unsuitable for mobile communications until recently, large amounts of new bandwidth are now becoming available.
\item Employing more base-station (BS) antennas: The spectral efficiency can be greatly improved by increasing the spatial dimensions according to the Multiple-Input-Multiple-Output (MIMO) theory \cite{Telatar,Foschini}. Significant gains have been demonstrated by adopting a massive number of either co-located BS antennas \cite{Lozano,Marzetta,FRusek,JSAC_LargeMIMO} (popularly known as \emph{massive MIMO}) or distributed BS antennas \cite{Roh,HDai,Zhuang,Choi,JZhang1,Zhu,Lee,Wang}.
\end{enumerate}

Among the new solutions, the distributed antenna system (DAS) is gaining increasing momentum, and emerging as a highly promising candidate for the 5G mobile communication systems \cite{Hu,WirelessCom_DAS,JSAC_DAS,Heath_Overview}. In a DAS, many low-power remote antenna ports are geographically distributed over a large area and connected to a central processor by fiber. The appealing features of distributed antennas have attracted considerable attention from both industry and academia, and been applied to the cutting-edge technologies such as small cells\footnote{The cellular system with small cells can be regarded as a special case of DAS if replacing each remote antenna port by a mini BS and requiring that each mobile user only communicates to the closest mini BS.} and the Cloud Radio Access Network (C-RAN) \cite{CRAN}.

\subsection{BS Antennas: Distributed or Co-located}

Compared to antenna arrays, the use of distributed antennas provides a much more efficient utilization of spatial resources. Specifically, signals to/from distributed BS antennas are subject to independent and different levels of large-scale fading. In contrast to the co-located case where the channel from the BS antenna array to each user becomes increasingly deterministic as the number of BS antennas grows, the channel randomness caused by the small-scale fading can be always preserved even with a large amount of distributed antennas. As a result, if the channel state information (CSI) is available at the transmitter side, the enhanced channel fluctuations can be fully utilized to provide higher water-filling gains and multiuser diversity gains \cite{Dai_JSAC}.

More importantly, with distributed BS antennas, the minimum access distance of each user can be greatly reduced by increasing the density of BS antennas. As a result, different rate scaling laws have been observed in cellular systems with co-located BS antennas or uniformly distributed BS antennas \cite{Dai_TWireless,Zhiyang,Junyuan}. For instance, it was shown in \cite{Junyuan} that by assuming that the number of BS antennas and the number of users grow infinitely with a fixed ratio $\upsilon$, the asymptotic downlink average user rate with maximum ratio transmission (MRT) scales in the orders of $\log_2 \upsilon$ and $\frac{\alpha}{2}\log_2 \upsilon$ for the co-located and distributed BS antenna layouts, respectively, where $\alpha>2$ is the path-loss factor. The rate gains achieved by distributed BS antennas become even more prominent when an orthogonal precoding scheme such as zero-forcing beamforming (ZFBF) is adopted.

Note that the above results are obtained by assuming a coordinated transmission of all the BS antennas. In a large-scale DAS with hundreds (or even thousands) of geographically distributed BS antennas, the computational cost and CSI measurement overhead of such a joint processing would be prohibitively high. How to establish a scalable signal processing framework for DASs is a key challenge that needs to be addressed.

\subsection{Virtual-Cell based DAS}

To reduce the system complexity, the common practice of cellular networks is to divide a large area into a number of cells where a BS is placed at the center of each cell and serves users who fall into its coverage. As users are randomly located in each cell which have distinct access distances to their serving BS, there are always users at cell boundary areas who suffer from lower power efficiency and higher inter-cell interference. The performance disparity of users could even be exacerbated if distributed BS antennas are used in cellular networks, as the BS antennas at cell boundary areas suffer from significantly higher inter-cell interference than those at the cell center in the uplink \cite{Dai_TWireless}, and become strong interfering sources to the neighboring cell-edge users in the downlink \cite{Zhiyang}. As pointed out in \cite{Dai_TWireless}, the cell-edge problem has its roots in the cellular structure where cells are formed based on the coverage of each BS. Such a \emph{BS-centric} structure, nevertheless, cannot be justified when both users and BS antennas are scattered around. Instead, the signal processing may be performed based on the unit of ``virtual cells'' \cite{Dai_thesis,Dai_Globecom,Dai_CDMA}.

In a virtual-cell based DAS, each user chooses its own serving BS antenna set as its virtual cell. The reason why virtual cells are formed in a \emph{user-centric} manner is two-fold: 1) With a high density of distributed BS antennas, each user may usually find more than one BS antennas in its vicinity to communicate with; 2) The BS antenna set serving each user varies with the user's location. By choosing a few surrounding BS antennas as each user's virtual cell, the number of channels that need to be measured becomes small, leading to a much lower CSI measurement overhead compared to cellular systems with large antenna arrays. More importantly, by increasing the density of distributed BS antennas, both the average number of users who share their virtual cells and the average number of users served by each BS antenna decline \cite{Dai_TWireless}, indicating that the signal processing in a large-scale DAS can be performed in a local and scalable manner.

Note that a few concepts were proposed before which share similarities to virtual cell, but with remarkable differences. In cellular networks with coordinated multipoint (CoMP) transmission, for instance, clusters of geographically distributed BSs are also formed to serve the cell-edge users. In most studies, however, the clustering is performed in a BS-centric manner with a fixed and a-priori BS coordination pattern \cite{MKarakayali,JZhang4,HDahrouj,SAkoum,KHuang}, which still leads to ``cluster-edge'' users \cite{JZhang4,KHuang}. Dynamic BS clustering was recently proposed in \cite{JZhang2,NLee} as a way of forming coordinated BS clusters based on users' locations. Yet different from the idea of virtual cell, it still follows the conventional cellular structure where each user is associated and served solely by its closest BS, though BSs coordinate their transmissions by choosing beamforming vectors to null out the intra-cluster interference.

Such a BS-centric clustering is also highly inefficient for a large-scale DAS with more BS antennas than users where some BS antennas may become redundant as no users are nearby. In this case, the BS antenna selection should be performed from users' perspective. Different optimization frameworks were recently established for a $K$-user DAS with $L\gg K$ distributed BS antennas to maximize the weighted sum rate \cite{ALiu} or average energy efficiency \cite{JJoung} by optimizing BS antenna selection, transmission power and precoding vectors. Instead of choosing BS antennas for each user to form a virtual cell, however, a joint antenna selection is performed for all the users.


\subsection{Effect of Virtual Cell Size}

For virtual-cell based DASs, the virtual cell size, i.e., how many BS antennas should be included into each user's virtual cell, is a key system parameter. In this paper, instead of formulating a joint optimization of virtual cell selection and precoding vectors which, similar to the problems considered in \cite{ALiu,JJoung}, would be intractable due to its non-convex and combinatorial nature, we consider a relatively simpler question: for given precoding scheme, what is the optimal virtual cell size?

Specifically, we focus on a large-scale downlink DAS where $K\gg 1$ users and $L\gg 1$ BS antennas are uniformly distributed in a certain area, and each user chooses $V$ closest BS antennas as its virtual cell. We aim to analyze the effect of the virtual cell size $V$ on the average user rate performance under two typical scenarios: 1) each user is served by its virtual cell independently of others, and 2) multiple users are grouped together with joint transmission from their virtual cells.

In the no-grouping case, MRT is assumed to be adopted in each user's virtual cell, and the corresponding achievable ergodic rate of each user is derived as an explicit function of the large-scale fading coefficients from all the users to their virtual cells. As the average user rate is determined by the joint probability density function (pdf) of the large-scale fading coefficients that is intractable when the number of BS antennas $L$ and the number of users $K$ are large, an upper-bound of the average user rate is obtained to analyze the effect of the virtual cell size $V$, based on which a rule of thumb for the optimal virtual cell size $V^*$ is further developed. It is shown that $V^*$ is solely determined by the ratio of the total number of BS antennas $L$ to the number of users $K$, which is much smaller than $L$ when $K$ is large. Simulation results verify that the rule of thumb provides a good estimation for the optimal virtual cell size to maximize the average user rate.

Due to no cooperation among virtual cells, users suffer from significant interference, which greatly degrades the rate performance in the first scenario. Therefore, we further consider grouping multiple users and adopt ZFBF to eliminate the intra-group interference. A novel user grouping algorithm is proposed, where users whose virtual cells overlap are grouped together. In contrast to the conventional BS-centric clustering where users located at the cluster edge still suffer from much degraded performance due to strong interference from BSs in neighboring clusters, with the proposed virtual-cell based user grouping, all the users achieve relatively uniform rate performance with the lowest rate significantly improved. It is further shown that different from the no-grouping case where a small virtual cell size is preferred, in this case, the virtual cell size $V$ determines a tradeoff between the average user rate and complexity: with a larger $V$, the average user rate can be improved owing to a lower interference level, yet more users need to be jointly served, leading to higher signal processing complexity.

The remainder of this paper is organized as follows. Section II introduces the system model. The rate analysis with MRT adopted in each user's virtual cell is presented in Section III. User grouping is further considered in Section IV, and concluding remarks are summarized in Section V.

Throughout this paper, the superscript $\dag$ and $T$ denote conjugate transpose and transpose, respectively. $\mathbb{E}[\cdot]$ denotes the expectation operator. $\lceil\cdot\rceil$ denotes the ceiling operator. $\|\mathbf{x}\|$ denotes the Euclidean norm of vector $\mathbf{x}$.  $\mathbf{I}_{K}$ denotes a $K\times K$ identity matrix. $\mathbf{1}_{1\times L} $ denotes a $1 \times L$ vector with all entries one. $\mathbf{0}_{L\times 1} $ denotes an $L \times 1$ vector with all entries zero. $x\sim \mathcal{CN}(u,\sigma^2)$ denotes a complex Gaussian random variable with mean $u$ and variance $\sigma^2$. $|\it{\mathcal{X}}|$ denotes the cardinality of set $\it{\mathcal{X}}$.
\begin{figure}[t]
\begin{center}
\includegraphics[width=0.35\textwidth]{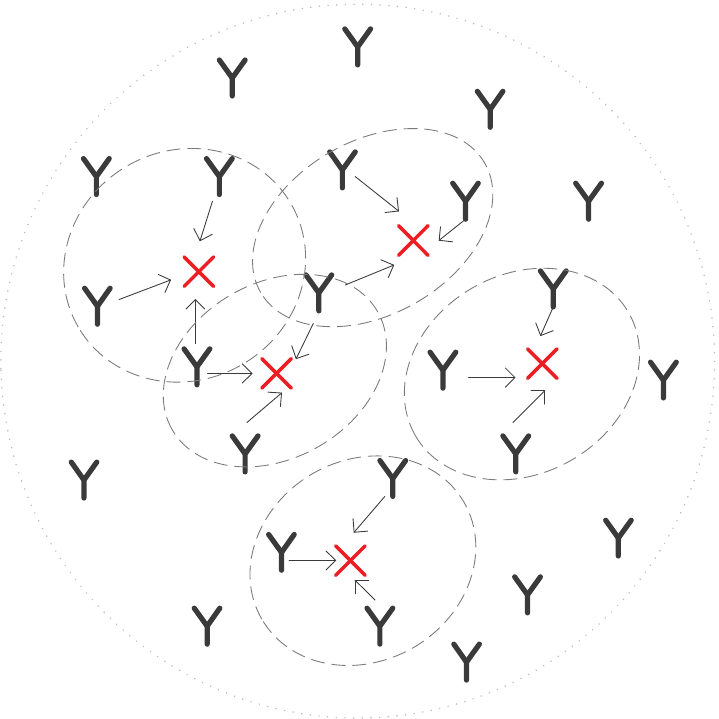}
\caption{Illustration of a downlink virtual-cell based DAS. $\times$ represents a user and Y represents a distributed BS antenna. Each user selects $V$ closest BS antennas to form its virtual cell.}
\label{FIG_DAS_Illustration}
\end{center}
\end{figure}

\section{System Model}
Consider a downlink virtual-cell based DAS with $K$ users and $L$ BS antennas uniformly distributed in a circular area with unit radius. Denote the set of users as $\mathcal{K}$ and the set of base-station (BS) antennas as $\mathcal{B}$, with $|\mathcal{K}|=K$ and $|\mathcal{B}|=L$. Assume that each user is equipped with a single antenna. As Fig. \ref{FIG_DAS_Illustration} illustrates, each user selects $V$ closest BS antennas to form its virtual cell.\footnote{Note that in general, each user may choose a different number of BS antennas to form its virtual cell. In this paper, we aim to analyze the effect of virtual cell size on the rate performance. Therefore, a homogeneous scenario where all the users have the same virtual cell size is considered.} For user $k\in\mathcal{K}$, denote its virtual cell as $\mathcal{V}_{k}$ with $|\mathcal{V}_{k}|=V$.

Let us first assume that each user is served by its virtual cell independently of others. This assumption will be relaxed in Section IV, where multiple users are grouped together and jointly served by their virtual cells. Let us focus on the downlink performance of user $k$. The received signal of user $k$ can be written as
\begin{equation}\label{signal model_NG}
y_{k}=\mathbf{g}_{k,\mathcal{V}_{k}}\mathbf{x}_{k,\mathcal{V}_{k}}
+\sum_{j\in \mathcal{K}, j\neq k}\mathbf{g}_{k,\mathcal{V}_{j}}\mathbf{x}_{j, \mathcal{V}_{j}}+n_{k},
\end{equation}
where $\mathbf{x}_{k,\mathcal{V}_{k}}\in\mathbb{C}^{V\times 1}$ is the transmitted signal vector from the virtual cell  $\mathcal{V}_{k}$ to user $k\in\mathcal{K}$.  $n_{k}\sim \mathcal{CN}(0,N_0)$ is the additive white Gaussian noise (AWGN) at user $k$. $\mathbf{g}_{k,\mathcal{V}_{k}}{\in}\mathbb{C}^{1\times V}$ denotes the channel gain vector from the virtual cell $\mathcal{V}_{k}$ to user $k$, which can be written as
\begin{equation}\label{define of g}
\mathbf{g}_{k,\mathcal{V}_{k}}=\bm{\gamma}_{k,\mathcal{V}_{k}}\circ\mathbf{h}_{k,\mathcal{V}_{k}},
\end{equation}
where $\mathbf{h}_{k,\mathcal{V}_{k}}\in\mathbb{C}^{1\times V}$ denotes the small-scale fading vector with entries modeled as independent and identically distributed (i.i.d) complex Gaussian random variables with zero mean and unit variance. $\bm{\gamma}_{k,\mathcal{V}_{k}}\in\mathbb{R}^{1\times V}$ is the large-scale fading vector from the virtual cell $\mathcal{V}_{k}$ to user $k$.  Without loss of generality, we ignore the shadowing effect and model the large-scale fading coefficient as
\begin{equation}\label{define of gamma}
\gamma_{k,l}=\|\mathbf{r}_{l}^{B}-\mathbf{r}_{k}^{U}\|^{-\frac{\alpha}{2}},
\end{equation}
where $\alpha$ is the path-loss factor. $\mathbf{r}_{l}^{B}$ is the position of the BS antenna $l\in\mathcal{B}$ and $\mathbf{r}_{k}^{U}$ is the position of user $k\in\mathcal{K}$. $\circ$ represents the Hadamard product.

Moreover, we assume that full channel state information (CSI) is perfectly known at both the transmitter side and the receiver side. With linear precoding, the transmitted signal from the virtual cell $\mathcal{V}_{k}$ to user $k$ can be written as
\begin{equation}\label{define of x}
\mathbf{x}_{k,\mathcal{V}_{k}}=\mathbf{w}_{k,\mathcal{V}_{k}} \cdot s_{k},
\end{equation}
for any $k\in\mathcal{K}$, where $s_{k}\sim \mathcal{CN}(0, \bar{P})$ is the information-bearing signal and $\mathbf{w}_{k,\mathcal{V}_{k}}$ is the precoding vector with $\|\mathbf{w}_{k,\mathcal{V}_{k}}\|=1$. The total transmission power for each user is assumed to be fixed at $\bar{P}$.

The second term on the right-hand side of (\ref{signal model_NG}), i.e, $u_{k}=\sum_{j\in \mathcal{K}, j\neq k}\mathbf{g}_{k,\mathcal{V}_{j}}\mathbf{x}_{j, \mathcal{V}_{j}}$, is the interference received at user $k$. With a large number of users and BS antennas, $u_{k}$ can be modeled as a complex Gaussian random variable with zero mean and variance $I_{k}$, which can be easily obtained from (\ref{define of x}) as
\begin{equation}\label{define of I_NG}
\begin{aligned}
I_{k}&=\sum_{j\in\mathcal{K}, j\neq k}
\mathbb{E}\left[\mathbf{g}_{k,\mathcal{V}_{j}}\mathbf{w}_{j,\mathcal{V}_{j}}
\mathbf{w}_{j,\mathcal{V}_{j}}^{\dag}\mathbf{g}_{k,\mathcal{V}_{j}}^{\dag}\right] \bar{P}
=\sum_{j\in\mathcal{K},j\neq k}\sum_{l\in\mathcal{V}_{j}}a_{j,l}\cdot \gamma_{k,l}^{2}\cdot \bar{P},
\end{aligned}
\end{equation}
where
\begin{equation}\label{define of al_NG}
\begin{aligned}
a_{j,l}&=\mathbb{E}_{\mathbf{h}_{j,\mathcal{V}_{j}}}\left[|w_{j,l}|^{2}\right],
\end{aligned}
\end{equation}
with $\sum_{l\in\mathcal{V}_{j}}a_{j,l}=1$ for any $j\in\mathcal{K}$.

In this paper, we normalize the total system bandwidth into unity and focus on the spectral efficiency. The achievable ergodic rate of user $k\in\mathcal{K}$ can be then written as
\begin{equation}\label{define of Rk_NG}
R_{k}=\mathbb{E}_{\mathbf{H}}\left[\log_{2}\left(1+
\frac{\bar{P}{\mathbf{g}}_{k,\mathcal{V}_{k}}\mathbf{w}_{k,\mathcal{V}_{k}}
\mathbf{w}_{k,\mathcal{V}_{k}}^{\dag}{\mathbf{g}}_{k,\mathcal{V}_{k}}^{\dag}}
{N_{0}+I_{k}}\right)\right],
\end{equation}
where the expectation is taken over the small-scale fading matrix $\mathbf{H}={[\mathbf{h}_{1, \mathcal{V}_{1}}^{T}, \mathbf{h}_{2, \mathcal{V}_{2}}^{T}, \cdots, \mathbf{h}_{K, \mathcal{V}_{K}}^{T}]}^{T}$.
Let
\begin{equation}\label{define of mu_NG}
\mu_{k}=\frac{\bar{P}\|\bm{\gamma}_{k, \mathcal{V}_{k}}\|^2}{N_{0}+I_{k}}
\end{equation}
denote the average received signal-to-interference-plus-noise ratio (SINR) of user $k$. By substituting (\ref{define of mu_NG}) into (\ref{define of Rk_NG}), the achievable ergodic rate $R_{k}$ can be further written as
\begin{equation}\label{Rk_mu_NG}
R_{k}=\mathbb{E}_{\mathbf{H}}\left[\log_{2}\left(1+\mu_{k}
\tilde{\mathbf{g}}_{k,\mathcal{V}_{k}}\mathbf{w}_{k,\mathcal{V}_{k}}
\mathbf{w}_{k,\mathcal{V}_{k}}^{\dag}\tilde{\mathbf{g}}_{k,\mathcal{V}_{k}}^{\dag}\right)\right],
\end{equation}
where
\begin{equation}\label{define of NormalizedG_NG}
\tilde{\mathbf{g}}_{k,\mathcal{V}_{k}}=\bm{\beta}_{k,\mathcal{V}_{k}}\circ \mathbf{h}_{k,\mathcal{V}_{k}}
\end{equation}
denotes the normalized channel gain vector from the virtual cell $\mathcal{V}_{k}$ to user $k$, and $\bm{\beta}_{k,\mathcal{V}_{k}}$ is the normalized large-scale fading vector with entries
\begin{equation}\label{define of beta_NG}
\beta_{k,l}=\frac{\gamma_{k,l}}{\|\bm{\gamma}_{k,\mathcal{V}_{k}}\|},
\end{equation}
$l\in\mathcal{V}_{k}$. Obviously, we have $\sum_{l\in\mathcal{V}_{k}}\beta_{k,l}^{2}=1$ for any $k\in\mathcal{K}$.

It can be clearly seen from (\ref{Rk_mu_NG}) that the rate performance is closely dependent on the precoding scheme adopted by each user's virtual cell. In the next section, we will take the example of MRT to demonstrate the effect of virtual cell size $V$ on the rate performance.

\section{Rate Performance without User Grouping}

In the single-user case, it is well known that MRT is optimal for achieving the system capacity \cite{Tse}. In this section, we assume that each user is served by its virtual cell using MRT. For any user $k\in\mathcal{K}$, the precoding vector is given by 
\begin{equation}\label{define of w_M}
\mathbf{w}_{k,\mathcal{V}_{k}}=\frac{\mathbf{g}_{k,\mathcal{V}_{k}}^{\dag}}{\|\mathbf{g}_{k,\mathcal{V}_{k}}\|}.
\end{equation}
In the following, we will first derive the achievable ergodic rate of each user as an explicit function of the large-scale fading coefficients.

\subsection{Achievable Ergodic Rate $R_{k}$ with MRT}
By substituting (\ref{define of w_M}) into (\ref{Rk_mu_NG}), the achievable ergodic rate of user $k$ with MRT can be obtained as
\begin{equation}\label{Rk_mu_NG_M}
R_{k}=\mathbb{E}_{\mathbf{h}_{k,\mathcal{V}_{k}}}
\left[\log_{2}\left(1+\mu_{k}\|\tilde{\mathbf{g}}_{k,\mathcal{V}_{k}}\|^{2}\right)\right],
\end{equation}
where the average received SINR $\mu_{k}$ can be obtained by combining (\ref{define of I_NG}), (\ref{define of mu_NG}) and (\ref{define of w_M}) as
\begin{equation}\label{mu_NG}
\mu_{k}=\frac{\|\bm{\gamma}_{k,\mathcal{V}_{k}}\|^2}
{\frac{N_{0}}{\bar{P}}+\sum_{j\in\mathcal{K},j\neq k}\sum_{l\in\mathcal{V}_{j}}
\Upsilon\left(\gamma_{j,l};\left\{\gamma_{j,i}\right\}_{i\neq l, i\in\mathcal{V}_{j}}\right)\cdot \gamma_{k,l}^{2}},
\end{equation}
with $\Upsilon\left(x; b_{1}, b_{2}, \cdots, b_{V-1}\right)$ defined as
\begin{align}\label{A}
\Upsilon\left(x; b_{1}, b_{2}, \cdots, b_{V-1}\right)
&=\left\{ {\begin{array}{*{20}{c}}
{{1}}&{\text{if}\quad V=1,}\\
{\sum_{m=1}^{V-1}
\frac{x^{-2}b_{m}^{-2}\left(\log x^{-2}-\log b_{m}^{-2}-1\right)+b_{m}^{-4}}
{\left(x^{-2}-b_{m}^{-2}\right)^{2}}
\prod_{t=1, t\neq m}^{V-1}\frac{b_{t}^{-2}}{b_{t}^{-2}-b_{m}^{-2}}}&{\text{otherwise}.}
\end{array}} \right.
\end{align}
Appendix \ref{Ap_mu} presents the detailed derivation of (\ref{mu_NG}).

As the normalized channel gain $\|\tilde{\mathbf{g}}_{k,\mathcal{V}_{k}}\|^{2}$ is a hypoexponential random variable with the corresponding probability density function (pdf) given by
\begin{equation}\label{pdf of NormaliedG}
\begin{aligned}
f_{\|\tilde{\mathbf{g}}_{k,\mathcal{V}_{k}}\|^{2}}(x)
=\sum_{l\in\mathcal{V}_{k}}\beta_{k,l}^{-2}\exp\left\{-\beta_{k,l}^{-2}x\right\}
\prod_{q\in\mathcal{V}_{k}, q\neq l}\frac{\beta_{k,q}^{-2}}{\beta_{k,q}^{-2}{-}\beta_{k,l}^{-2}},
\end{aligned}
\end{equation}
by combining (\ref{define of beta_NG}) and (\ref{Rk_mu_NG_M}-\ref{pdf of NormaliedG}), the achievable ergodic rate of user $k$ can be finally obtained as
\begin{equation}\label{Rk_NG}
\begin{aligned}
R_{k}=&\sum_{l\in\mathcal{V}_{k}}\exp\left\{\frac{\frac{N_{0}}{\bar{P}}+\sum_{j\in\mathcal{K},j\neq k}\sum_{m\in\mathcal{V}_{j}}
\Upsilon\left(\gamma_{j,m};\left\{\gamma_{j,i}\right\}_{i\neq m, i\in\mathcal{V}_{j}}\right)\cdot\gamma_{k,m}^{2}}
{\gamma_{k,l}^{2}}\right\}\cdot \\
&{E_{1}\left\{\frac{\frac{N_{0}}{\bar{P}}+\sum_{j\in\mathcal{K},j\neq k}\sum_{m\in\mathcal{V}_{j}}
\Upsilon\left(\gamma_{j,m};\left\{\gamma_{j,i}\right\}_{i\neq m, i\in\mathcal{V}_{j}}\right)\cdot \gamma_{k,m}^{2}}
{\gamma_{k,l}^{2}}\right\}\prod_{q\in\mathcal{V}_{k}, q\neq l}\frac{\gamma_{k,q}^{-2}}{\gamma_{k,q}^{-2}{-}\gamma_{k,l}^{-2}}\log_{2}e},
\end{aligned}
\end{equation}
where $E_{1}\left\{x\right\}=\int_{x}^{\infty}t^{-1}e^{-t}dt$.

\begin{figure*}[t]
\centering
{\subfloat[]{\includegraphics[width=3.1in]{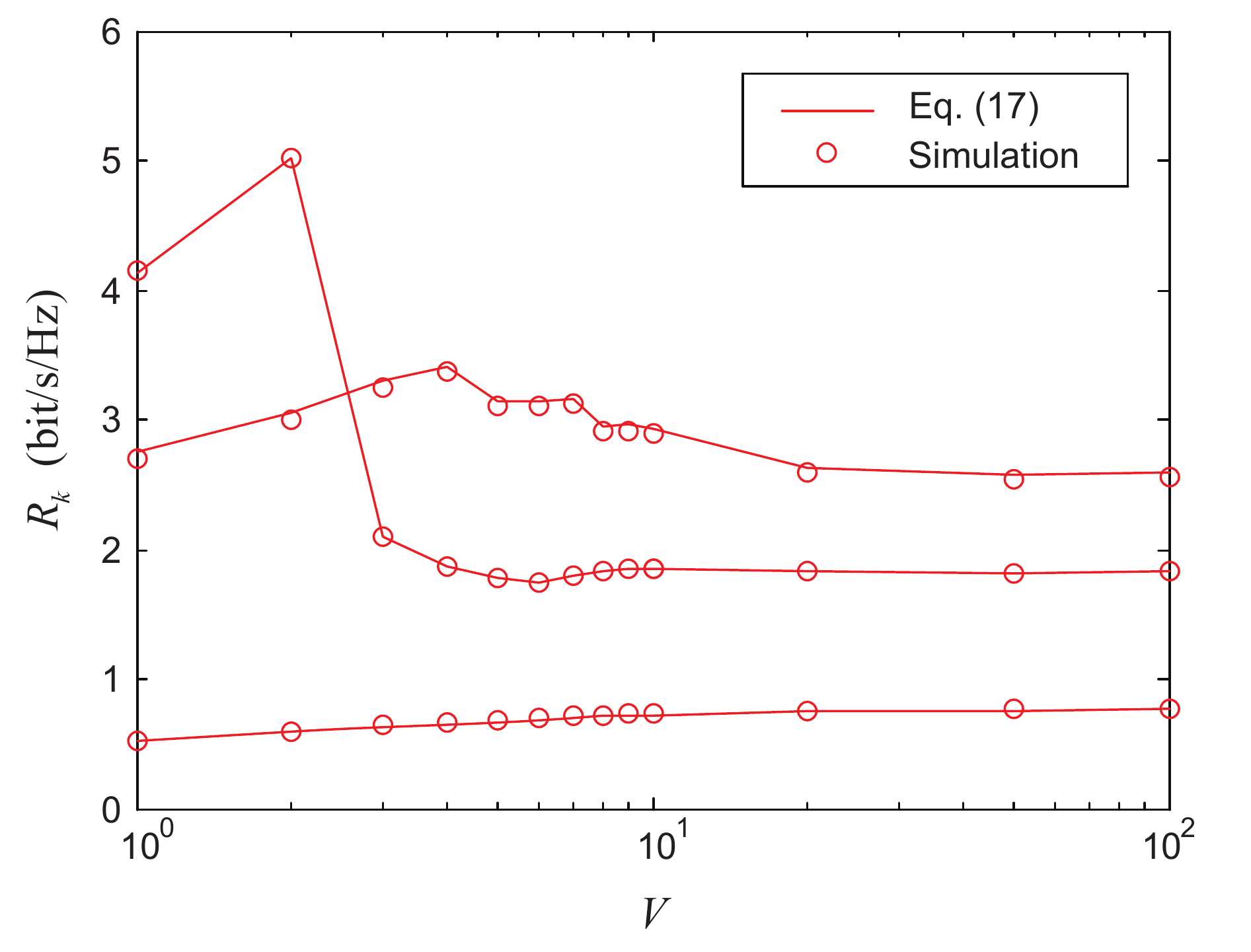}
\label{FIG_R_NG_K50L100_FixedL}}\hfil
\subfloat[]{\includegraphics[width=3.1in]{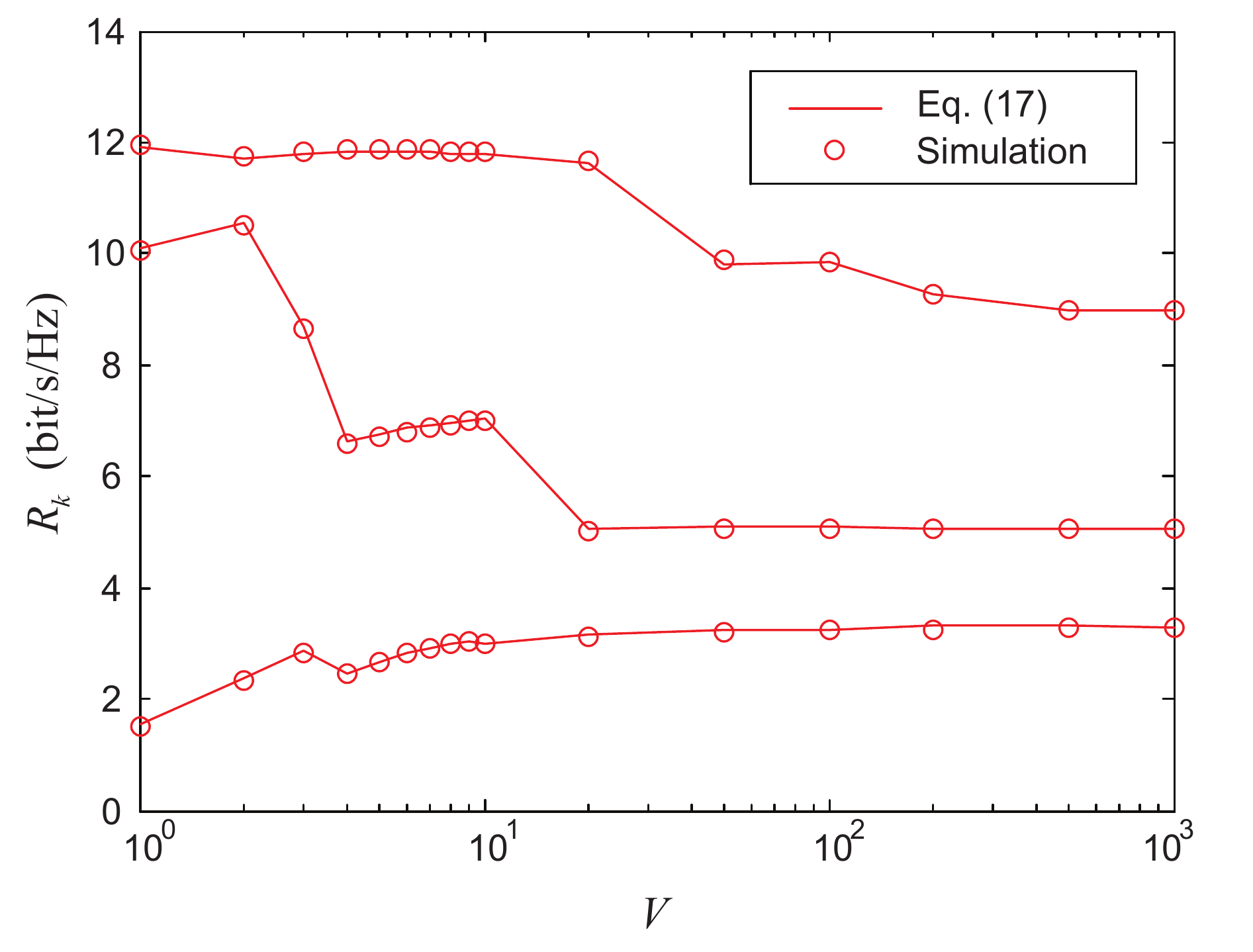}
\label{FIG_R_NG_K50L1000_FixedL}}}
\caption{Achievable ergodic rate $R_{k}$ of a randomly selected user $k$ versus virtual cell size $V$ under 3 realizations of the large-scale fading coefficients $\{\gamma_{j,l}\}_{j\in\mathcal{K},l\in\mathcal{B}}$. $\alpha=4$. $\bar{P}/N_{0}=10$dB. $K=50$. (a) $L=100$. (b) $L=1000$.} 
\label{FIG_R_NG}
\end{figure*}

It is clear from (\ref{Rk_NG}) that the achievable ergodic rate of user $k$ is a function of the large-scale fading coefficients of all the users, i.e,  $\{\gamma_{j,l}\}_{j\in\mathcal{K},l\in\mathcal{B}}$. Fig. \ref{FIG_R_NG} presents the simulation results of the achievable ergodic rate $R_{k}$ of a randomly selected user $k$ given 3 random realizations of the large-scale fading coefficients\footnote{More specifically, for each realization, we generate the positions of $K$ users and $L$ BS antennas which are supposed to be uniformly distributed in a circular area with unit radius, and then calculate the large-scale fading coefficients based on (\ref{define of gamma}).} $\{\gamma_{j,l}\}_{j\in\mathcal{K},l\in\mathcal{B}}$ with the total number of BS antennas $L=100$ and $1000$, and the total number of users $K=50$. It can be clearly observed from Fig. \ref{FIG_R_NG} that (\ref{Rk_NG}) is accurate and the rate performance of user $k$ is sensitive to the large-scale fading coefficients $\{\gamma_{j,l}\}_{j\in\mathcal{K},l\in\mathcal{B}}$. Moreover, by comparing Fig. \ref{FIG_R_NG_K50L100_FixedL} and Fig. \ref{FIG_R_NG_K50L1000_FixedL}, we can see that on average, the achievable ergodic rate can be significantly improved by increasing the total number of BS antennas $L$. Yet the effect of the virtual cell size $V$ on the achievable ergodic rate depends on the specific realization of the large-scale fading coefficients $\{\gamma_{j,l}\}_{j\in\mathcal{K},l\in\mathcal{B}}$.

For networks with a large number of users, the \textit{average} rate performance per user is an important performance metric. In our case, as the achievable ergodic rate of each user is closely dependent on the large-scale fading coefficients $\{\gamma_{j,l}\}_{j\in\mathcal{K},l\in\mathcal{B}}$, in the following, we would focus on the average achievable ergodic rate (which is referred to as ``average user rate'') defined by
\begin{equation}\label{define of AverageR_NG}
\bar{R}\triangleq\mathbb{E}_{\{\gamma_{j,l}\}_{j\in\mathcal{K}, l\in\mathcal{B}}}\left[R_{k}\right].
\end{equation}
We are specifically interested in the effect of the virtual cell size $V$ on the average user rate $\bar{R}$.

\subsection{Effect of Virtual Cell Size $V$ on Average User Rate $\bar{R}$}
The achievable ergodic rate $R_{k}$ of user $k$ is derived in (\ref{Rk_NG}) as an explicit function of the large-scale fading coefficients of all the users $\{\gamma_{j,l}\}_{j\in\mathcal{K},l\in\mathcal{B}}$. With BS antennas uniformly distributed within a circular area, the pdf of the large-scale fading coefficient from a given user to a BS antenna has been derived in \cite{Dai_JSAC}. Nevertheless, it is difficult to obtain the joint pdf of all the large-scale fading coefficients $\{\gamma_{j,l}\}_{j\in\mathcal{K},l\in\mathcal{B}}$ especially when the number of BS antennas $L$  and  the number of users $K$ are large. Therefore, we resort to an upper-bound to study the effect of the virtual cell size $V$ on the average user rate $\bar{R}$ in the following.

\subsubsection{An Upper-bound of Average User Rate $\bar{R}^{ub}$}
According to Jensen's inequality, the achievable ergodic rate of user $k$ given in (\ref{Rk_mu_NG_M}) is upper-bounded by
\begin{align}\label{Rk_mu_NG_ub}
R_{k}\leq \log_{2}\left(1+\mu_{k}\right).
\end{align}
Then the average user rate $\bar{R}$ defined in (\ref{define of AverageR_NG}) is upper-bound by
\begin{align}\label{AverageR_mu_NG_ub}
\bar{R}\leq
\mathbb{E}_{\{\gamma_{j,l}\}_{j\in\mathcal{K}, l\in\mathcal{B}}}\left[\log_{2}\left(1+\mu_{k}\right)\right].
\end{align}

It is clear from (\ref{mu_NG}) that the average received SINR $\mu_{k}$ is determined by two parts: (a) the normalized average received signal power
\begin{equation}\label{NormalizedSignal}
\tilde{S}_{k}=\|\bm{\gamma}_{k,\mathcal{V}_{k}}\|^2,
\end{equation}
and (b) the normalized interference power
\begin{equation}\label{NormalizedI}
\tilde{I}_{k}=\sum_{j\in\mathcal{K},j\neq k}\sum_{l\in\mathcal{V}_{j}}\Upsilon\left(\gamma_{j,l};\left\{\gamma_{j,i}\right\}_{i\neq l, i\in\mathcal{V}_{j}}\right)\cdot \gamma_{k,l}^{2},
\end{equation}
which is lower-bounded by
\begin{equation}\label{NormalizedI_lb}
\tilde{I}_{k}\geq \tilde{I}_{k}^{lb}=\sum_{l\in\mathcal{V}_{j^{*}}}\Upsilon\left(\gamma_{j^{*},l};\left\{\gamma_{j^{*},i}\right\}_{i\neq l, i\in\mathcal{V}_{j^{*}}}\right)\cdot \gamma_{k,l}^{2},
\end{equation}
where $j^{*}$ denotes the closest interfering user of user $k$. An upper-bound of the average user rate can be then obtained by combining (\ref{mu_NG}) and (\ref{AverageR_mu_NG_ub}-\ref{NormalizedI_lb}) as
\begin{align}\label{AverageR_mu_ub_NG_ub}
\bar{R}\leq \bar{R}^{ub}= \mathbb{E}_{\{\gamma_{j,l}\}_{j\in\mathcal{K}, l\in\mathcal{B}}}
\left[\log_{2}\left(1+\frac{\tilde{S}_{k}}{\tilde{I}_{k}^{lb}}\right)\right].
\end{align}
Appendix \ref{Ap_LogSI} further shows that the upper-bound $\bar{R}^{ub}$ can be obtained as
\begin{align}\label{AverageR_NG_ub}
\bar{R}^{ub}
=&\frac{2^{V+1}}{\pi^{V}}(K-1)\left(\frac{L!}{(L-V)!}\right)^2
\int_{0}^{1}z\left(1-z^{2}\right)^{K-2}
\underbrace{\int_{0}^{1}\int_{0}^{y_{V}}\int_{0}^{y_{V-1}}\cdots\int_{0}^{y_{2}}}_{V-fold}
\left(1-y_{V}^{2}\right)^{L-V} \nonumber \\
&\prod_{i=1}^{V}y_{i}\underbrace{\int_{0}^{1}\int_{0}^{x_{V}}\int_{0}^{x_{V-1}}\cdots\int_{0}^{x_{2}}}_{V-fold}
\left(1-x_{V}^{2}\right)^{L-V}\prod_{i=1}^{V}x_{i}
\underbrace{\int_{0}^{2\pi}\int_{0}^{2\pi}\cdots\int_{0}^{2\pi}}_{V-fold} \nonumber \\
&\log_{2}\left(1+\frac{\sum_{i=1}^{V}x_{i}^{-\alpha}}
{\sum_{i=1}^{V}\Upsilon\left(y_{i}^{-\frac{\alpha}{2}}; y_{1}^{-\frac{\alpha}{2}}, \cdots, y_{i-1}^{-\frac{\alpha}{2}}, y_{i+1}^{-\frac{\alpha}{2}}, \cdots, y_{V}^{-\frac{\alpha}{2}} \right)\cdot
\left(y_{i}^{2}+z^{2}+2y_{i}z\cos\omega_{i}\right)^{-\frac{\alpha}{2}}}\right) \nonumber \\
&d\omega_{1}d\omega_{2}\cdots d\omega_{V}dx_{1}dx_{2}\cdots dx_{V}dy_{1}dy_{2}\cdots dy_{V}dz,
\end{align}
where $\Upsilon(x; b_{1}, b_{2}, \cdots, b_{V-1})$ is given in (\ref{A}).

\begin{figure*}[t]
\centering
{\subfloat[]{\includegraphics[width=3in]{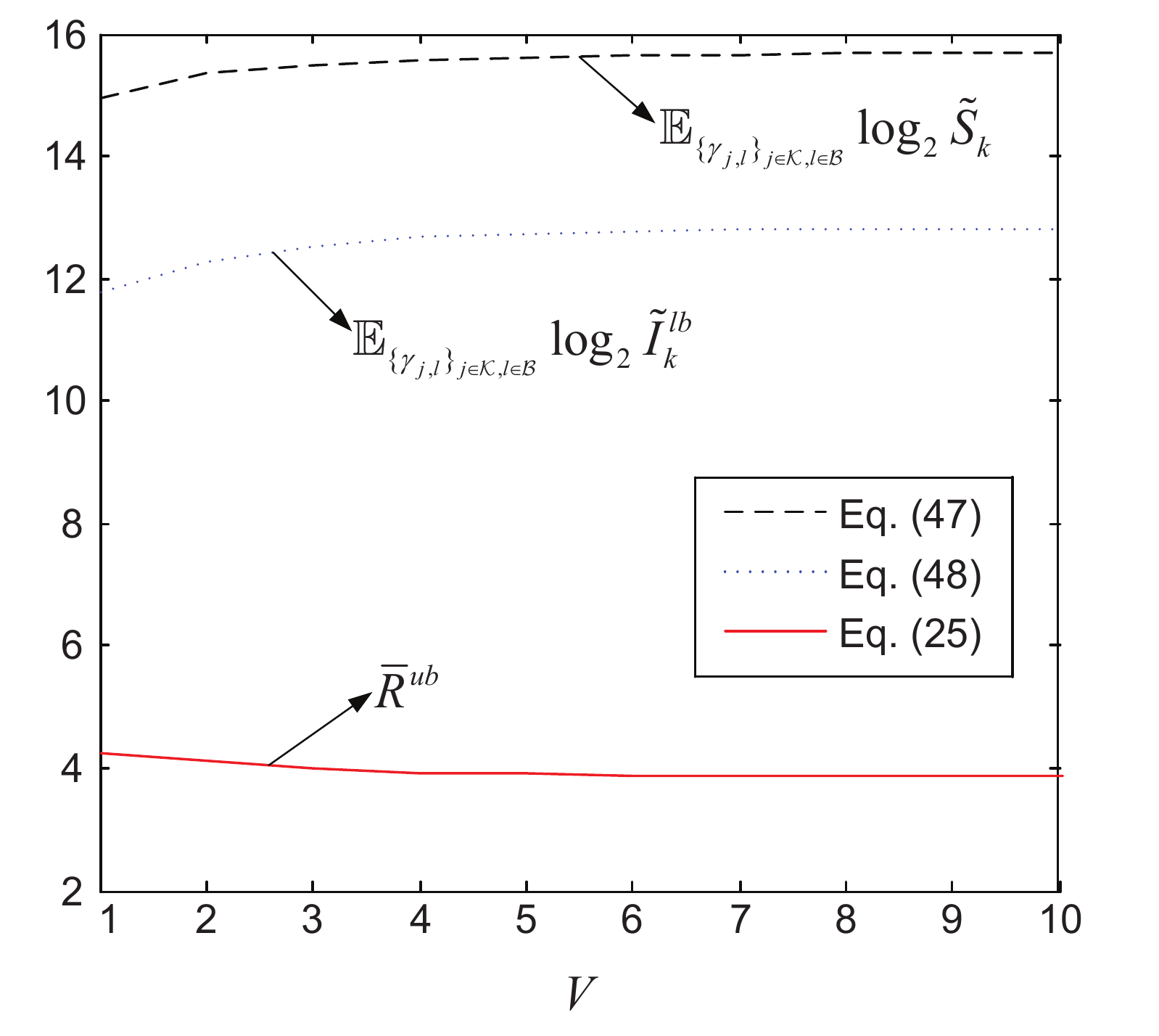}
\label{FIG_ALog_K50L100}}\hfil
\subfloat[]{\includegraphics[width=3in]{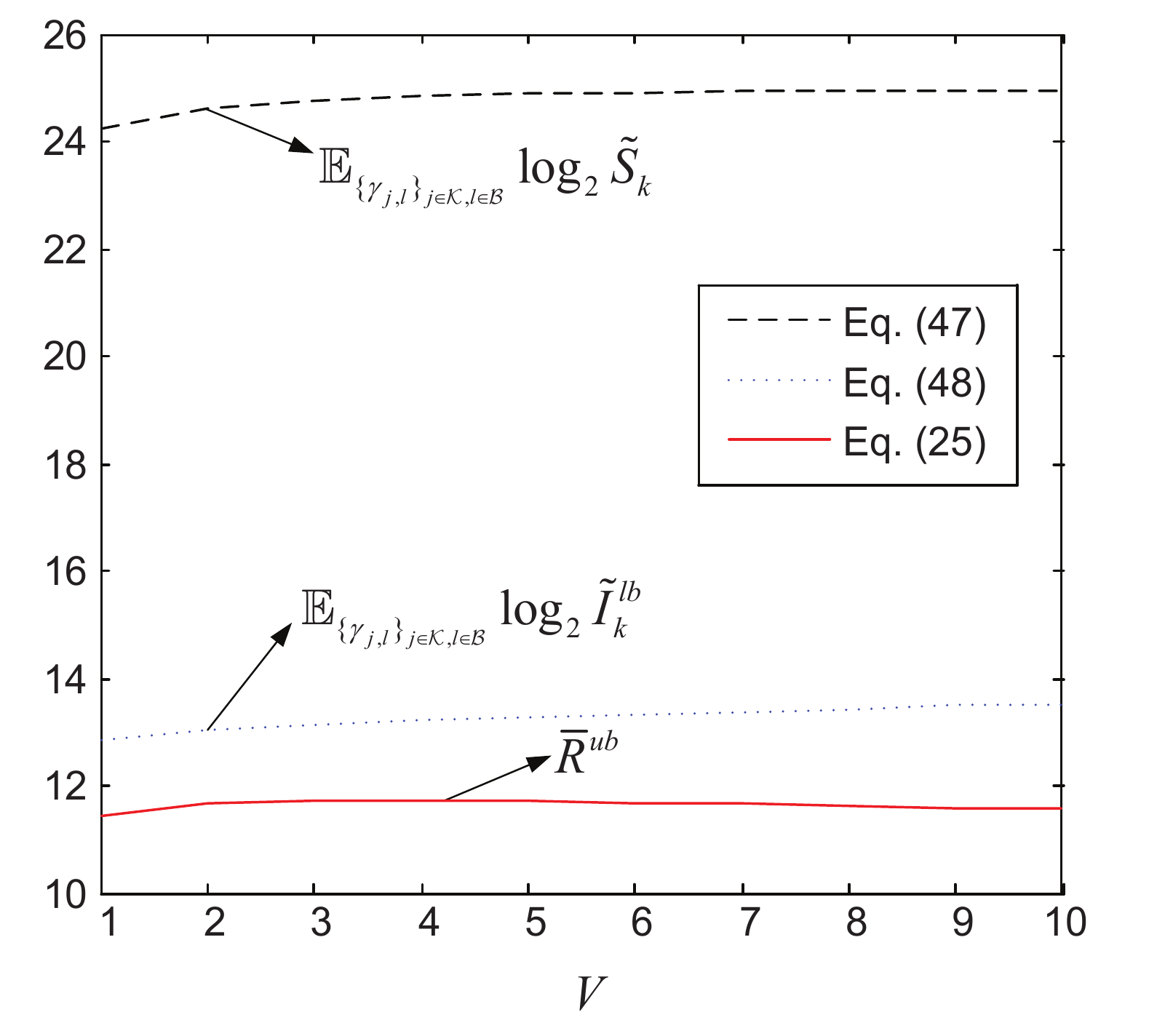}
\label{FIG_ALog_K50L2500}}}
\caption{Upper-bound of the average user rate $\bar{R}^{ub}$ versus virtual cell size $V$. (a) $L=100$. (b) $L=2500$. $\alpha=4$. $\bar{P}/N_{0}=10$dB. $K=50$.} 
\label{FIG_NG_K50}
\end{figure*}

Fig. \ref{FIG_NG_K50} presents the upper-bound of the average user rate $\bar{R}^{ub}$ given in (\ref{AverageR_NG_ub}) under various values\footnote{Note that $\bar{R}^{ub}$ is derived in (\ref{AverageR_NG_ub}) as a $3V$-fold integral. Although the numerical calculation of multi-dimensional integration can be effectively performed by using numerical softwares such as Matlab, the computational complexity sharply increases with $V$. Therefore, we only present the results of $\bar{R}^{ub}$ for $V\le 10$.} of the virtual cell size $V$ with the number of users $K=50$ and the number of BS antennas $L=100$ and $2500$. For the sake of illustration, the average entropy of the normalized received signal power $\mathbb{E}_{\{\gamma_{j,l}\}_{j\in\mathcal{K}, l\in\mathcal{B}}}\log_{2}\tilde{S}_{k}$ and the average entropy of the normalized interference power lower-bound $\mathbb{E}_{\{\gamma_{j,l}\}_{j\in\mathcal{K}, l\in\mathcal{B}}}\log_{2}\tilde{I}_{k}^{lb}$, which are given in (\ref{E_log_S}) and (\ref{E_logI_lb}) in Appendix \ref{Ap_LogSI}, respectively, are also plotted in Fig. \ref{FIG_NG_K50}. It can be clearly seen from Fig. \ref{FIG_ALog_K50L100} that both $\mathbb{E}_{\{\gamma_{j,l}\}_{j\in\mathcal{K}, l\in\mathcal{B}}}\log_{2}\tilde{S}_{k}$ and $\mathbb{E}_{\{\gamma_{j,l}\}_{j\in\mathcal{K}, l\in\mathcal{B}}}\log_{2}\tilde{I}_{k}^{lb}$ increase with the virtual cell size $V$ and converge when $V$ is large. Intuitively, with a large virtual cell size $V$, more BS antennas are included for each user's signal transmission, and thus both the received signal power and the interference power are increased on average. When $V$ is large enough, however, the newly added BS antennas in each user's virtual cell are far away from the user with negligible contributions under MRT. As a result, the effect of $V$ on the rate performance becomes marginal. As we can see from Fig. \ref{FIG_ALog_K50L100}, the upper-bound of the average user rate significantly varies with the virtual cell size $V$ only when $V$ is small.

It can be also observed from Fig. \ref{FIG_NG_K50} that there exists an optimal virtual cell size $V^{*}$ to maximize the upper-bound of the average user rate, and $V^{*}$ depends on the total number of BS antennas $L$. Specifically, with $L=100$ in Fig. \ref{FIG_ALog_K50L100}, $\bar{R}^{ub}$ is maximized when $V=1$. When $L$ increases to $2500$, it can be seen from Fig. \ref{FIG_ALog_K50L2500} that the optimal virtual cell size $V^{*}$ becomes larger than $1$. As the optimization of $\bar{R}^{ub}$, which is obtained in (\ref{AverageR_NG_ub}) as a multi-fold integral, is intractable, we will develop a rule of thumb for choosing the optimal virtual cell size $V^{*}$ in the next subsection.

\subsubsection{A Rule of Thumb for Optimal Virtual Cell Size $V^{*}$}
\begin{figure}[t]
\begin{center}
\includegraphics[width=0.4\textwidth]{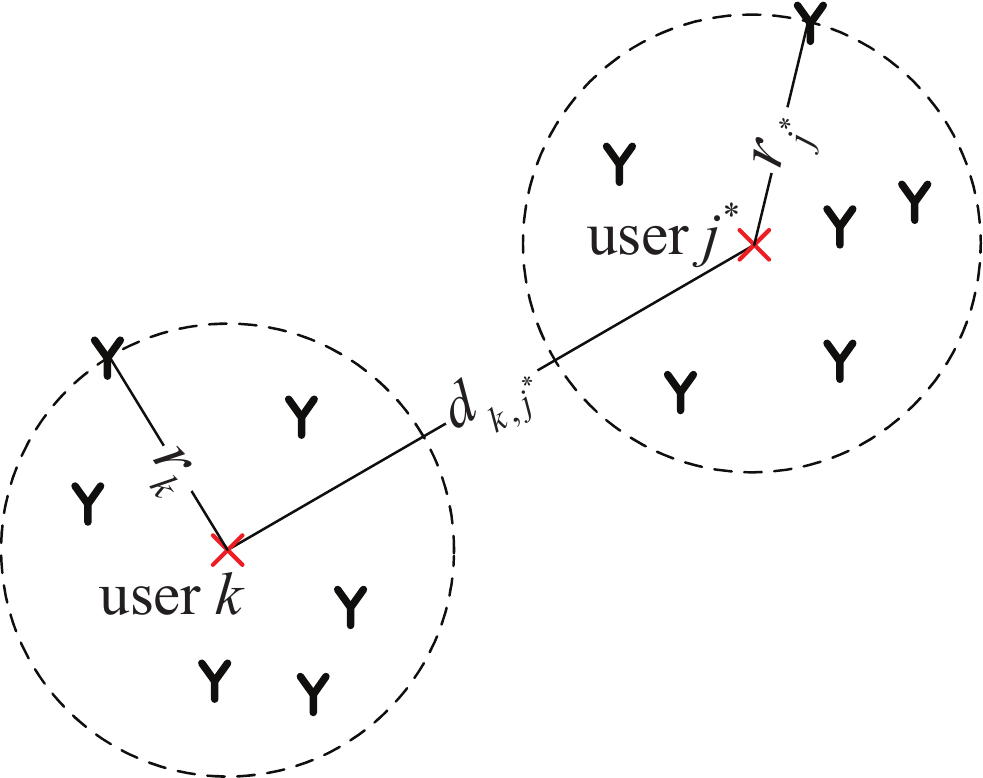}
\caption{Illustration of the radiuses of the virtual cells of user $k$ and its closest interfering user $j^{*}$. $\times$ represents a user and Y represents a distributed BS antenna.}
\label{FIG_OptV_Illustration}
\end{center}
\end{figure}

To visualize the effect of the virtual cell size $V$ on the rate performance, let us first define the radius of a user's virtual cell as the distance from this user to its $V$th closest BS antenna. Fig. \ref{FIG_OptV_Illustration} illustrates the radiuses of the virtual cells of user $k$ and its closest interfering user $j^{*}$.\footnote{Similar to the derivation of the upper-bound $\bar{R}^{ub}$, here we only consider user $k$'s closest interfering user $j^{*}$ whose signal contributes the most to the interference of user $k$.} Intuitively, as the virtual cell size $V$ increases, the radius of user $k$'s virtual cell $r_{k}$ becomes larger, indicating that more BS antennas are included into user $k$'s virtual cell and thus the average received signal power of user $k$ increases. On the other hand, with a larger $V$, the radius of user $j^{*}$'s virtual cell $r_{j^{*}}$ would also increase, with which more BS antennas transmit to user $k$'s closest interfering user $j^{*}$, leading to higher interference for user $k$, as Fig. \ref{FIG_NG_K50} shows. If $V$ is too large, the virtual cells of user $k$ and user $j^{*}$ would overlap with each other, in which case user $k$ would suffer from strong interference from user $j^{*}$ due to the shared BS antennas. Therefore, to reach a fine balance between the average received signal power and interference, we propose to choose the largest virtual cell size without causing an overlap of user $k$ and user $j^{*}$'s virtual cells. That is,
\begin{align}\label{Opt_Prob}
&\text{max} \qquad V \\ 
&\text{s.t.} \; \qquad r_{k}+r_{{j}^{*}}\leq d_{k,j^{*}},
\end{align}
where $d_{k,j^{*}}$ is the distance between user $k$ and user $j^{*}$. As we focus on the average user rate, we further relax the constraint to
\begin{align}\label{constraint_relax}
\mathbb{E}_{\{\gamma_{j,l}\}_{j\in\mathcal{K}, l\in\mathcal{B}}}\left[r_{k}+r_{{j}^{*}}\right]\leq \bar{d}_{k,j^{*}},
\end{align}
where $\bar{d}_{k,j^{*}}=\mathbb{E}_{\{\gamma_{j,l}\}_{j\in\mathcal{K}, l\in\mathcal{B}}}\left[{d}_{k,j^{*}}\right]$.

Appendix \ref{Ap_OptV} shows that the solution of the optimization problem defined in (\ref{Opt_Prob}) and (\ref{constraint_relax}) is given by
\begin{align}\label{OptV_NG}
V^{*}&=\frac{L}{4}
\left(\frac{K-1}{(K-2)^{\frac{3}{2}}}\left(-(K-2)^{\frac{1}{2}}e^{-(K-2)}+\frac{1}{2}\Gamma\left(\frac{1}{2}, 0\right)-\frac{1}{2}\Gamma\left(\frac{1}{2}, K-2\right)\right)\right)^{2} \nonumber \\
&\mathop{\approx}\limits^{\text{for large}\; K}0.2\frac{L}{K},
\end{align}
where $\Gamma(s, x)=\int_{x}^{\infty}t^{s-1}e^{-t}dt$. (\ref{OptV_NG}) indicates that the optimal virtual cell size $V^{*}$ is solely determined by the ratio of the number of BS antennas $L$ to the number of users $K$. Intuitively, with an increasing number of BS antennas $L$, more BS antennas should be included in user $k$'s virtual cell to improve the average received signal power. In contrast, if the number of users $K$ increases, the virtual cell size $V$ should be reduced to avoid the overlap of different virtual cells.
Note that the size of each virtual cell should be an integer and no smaller than $1$. Therefore, (\ref{OptV_NG}) should be further refined as
\begin{align}\label{OptV_NG_Integer}
V^{*}=\left\lceil0.2\frac{L}{K}\right\rceil.
\end{align}

\begin{figure}[t]
\begin{center}
\includegraphics[width=0.6\textwidth]{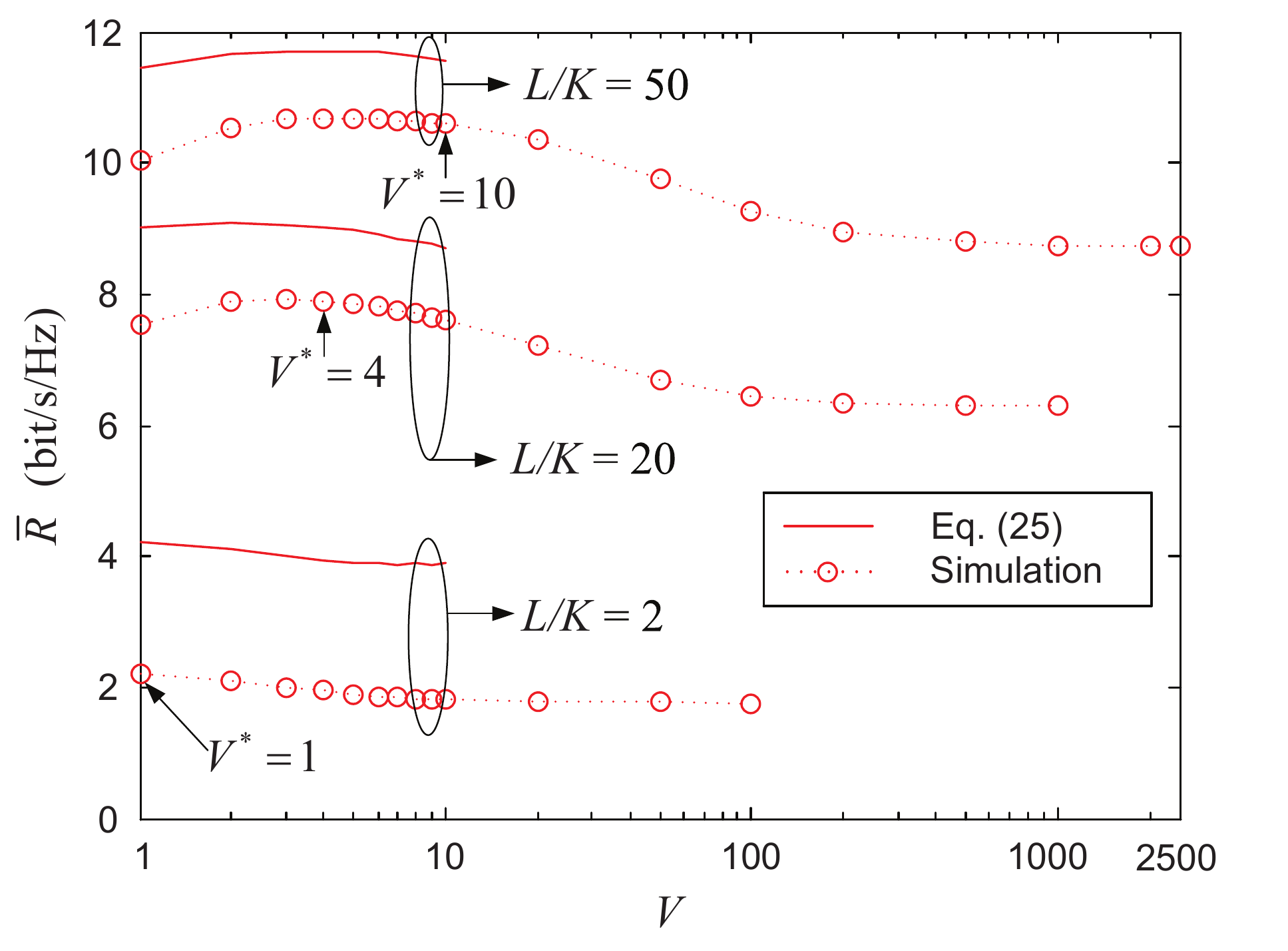}
\caption{Average user rate $\bar{R}$ versus virtual cell size $V$. For demonstration, the upper-bound of the average user rate $\bar{R}^{ub}$ given in (\ref{AverageR_NG_ub}) is also plotted. $\alpha=4$. $\bar{P}/N_{0}=10$dB. $K=50$. $L/K=2, 20, 50$.}
\label{FIG_OptV_NG}
\end{center}
\end{figure}

\subsubsection{Simulation Results}
Fig. \ref{FIG_OptV_NG} presents the simulation results of the average user rate $\bar{R}$ when the ratio of the number of BS antennas $L$ to the number of users $K$ is fixed at $2$, $20$ and $50$. For demonstration, the upper-bound of the average user rate $\bar{R}^{ub}$ given in (\ref{AverageR_NG_ub}) is also plotted. It can be clearly seen that $\bar{R}$ varies with the virtual cell size $V$ in the same manner as its upper-bound $\bar{R}^{ub}$. That is, it monotonically decreases as $V$ increases when the ratio $L/K=2$. If $L/K=20$ or $50$, in contrast, it first increases with $V$ and then decreases. It can be also observed from Fig. \ref{FIG_OptV_NG} that the rule of thumb developed in (\ref{OptV_NG_Integer}) provides a good estimation for the optimal virtual cell size. According to (\ref{OptV_NG_Integer}), the optimal virtual cell size $V^{*}$ needs to be enlarged as the ratio of the number of BS antennas $L$ to the number of users $K$ increases. Fig. \ref{FIG_OptV_NG} corroborates that as $L/K$ increases from $2$ to $50$, the optimal virtual cell size to maximize the average user rate should be increased from $1$ to $10$.

It can be also seen from (\ref{OptV_NG_Integer}) and Fig. \ref{FIG_OptV_NG} that with a large number of users, the optimal virtual cell size is much smaller than the total number of BS antennas $L$, which indicates that compared to a joint transmission from all the BS antennas, choosing a small number of adjacent BS antennas not only greatly lowers the complexity, but also leads to superior rate performance. Moreover, (\ref{OptV_NG_Integer}) shows that selecting the closest BS antenna, i.e., $V=1$, is optimal only when the ratio $L/K$ is small. For the future 5G mobile communication system which is expected to deploy a high density of BS antennas, multiple BS antennas would need to be selected to form each user's virtual cell.

\section{Rate Performance with User Grouping}

In Section III, we have analyzed the average user rate when no cooperation is adopted among different users' virtual cells. In that case, each user suffers from considerable interference, which greatly degrades the rate performance. In this section, we further consider grouping multiple users together with coordinated transmission from their virtual cells.

\begin{figure}[t]
\begin{center}
\includegraphics[width=2.7in]{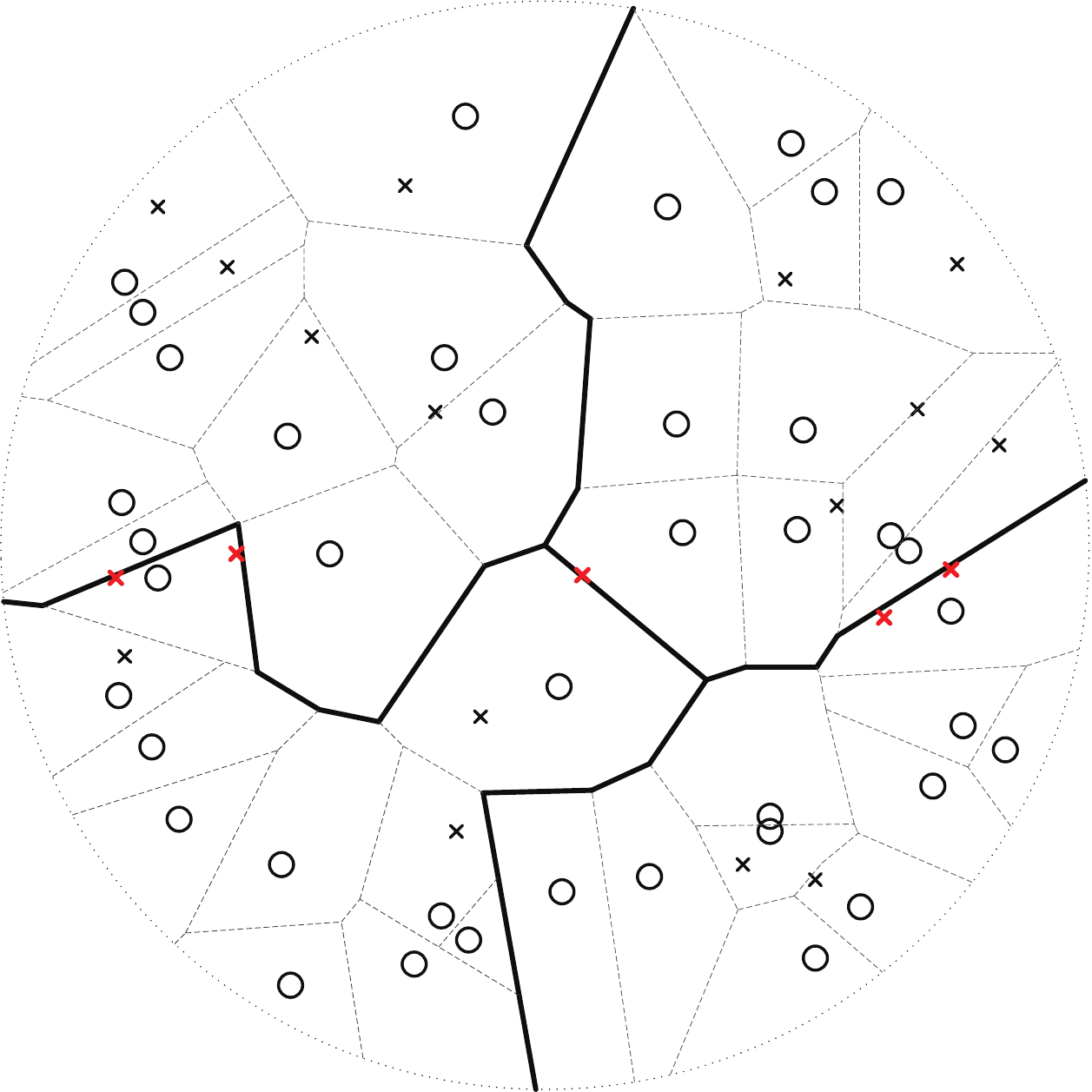}
\caption{Illustration of BS clustering. The BSs are divided into $4$ clusters with $10$ BSs in each cluster. $\times$ represents a user and $\circ$ represents a BS. The cell edge is plotted in dotted lines and the cluster edge is plotted in solid lines. $L=40$. $K=20$.}
\label{FIG_Snapshot_BS}
\end{center}
\end{figure}

\begin{algorithm}
\caption{Virtual-Cell based User Grouping}
\label{Alg_Overlap}
\begin{algorithmic}[1]
\State Initialization: Each user $k\in\mathcal{K}$ selects its $V$ closest BS antennas to form its virtual cell $\mathcal{V}_{k}$. $m=0$.
\While{$\mathcal{K}\neq \emptyset$}
\State $\forall$ any user $k\in \mathcal{K}$, set the user group $\mathcal{K}_{m}=\{k\}$ and the BS antenna set $\mathcal{B}_{m}=\mathcal{V}_{k}$. Delete user $k$ from $\mathcal{K}$.
\While{$\mathcal{B}_{m}\cap\left(\cup_{j\neq k, j\in\mathcal{K}}\mathcal{V}_{j}\right)\neq \emptyset$}
\For{$j\in\mathcal{K}$}
\If {$\mathcal{V}_{j}\cap\mathcal{B}_{m}\neq \emptyset$}
\State $j\in\mathcal{K}_{m}$ and $\mathcal{B}_{m}=\mathcal{B}_{m}\cup\mathcal{V}_{j}$. Delete user $j$ from $\mathcal{K}$.
\EndIf
\EndFor
\EndWhile
\State $m=m+1$.
\EndWhile
\end{algorithmic}
\end{algorithm}

\subsection{Virtual-Cell based User Grouping}

As we mentioned in Section I, for cellular systems with CoMP transmission, multiple BSs are normally clustered so as to improve the rate performance of cell-edge users. As we can see from Fig. \ref{FIG_Snapshot_BS}, in this case, users located at the \emph{cluster edge} still suffer from much degraded performance due to strong interference from BSs in neighboring clusters. The edge effect indeed originates from the BS-centric structure. As different users could be geographically close to different sets of BSs, clustering BSs would always lead to a few ``unlucky'' ones whose closest BSs do not belong to the same cluster and thus cannot coordinate their transmissions. To ensure that every user is served by its best BS set, the grouping should be user-centric, and the information of users' virtual cells should be fully utilized for user grouping.

More specifically, in a virtual-cell based DAS, since each user forms its virtual cell by selecting $V$ closest BS antennas, users who are close around tend to choose the same, or partially the same, BS antenna set as their virtual cells, and cause strong interference to each other if no coordination is performed among their virtual cells. Therefore, we propose the following virtual-cell based user grouping algorithm: For any users $k,j\in \mathcal{K}$, if their virtual cells overlap with each other, i.e., $\mathcal{V}_{k}\cap \mathcal{V}_{j}\neq \emptyset$, then group them together and merge their virtual cells. Repeat this step until the whole user set is divided into disjoint subsets. The detailed description is presented in Algorithm \ref{Alg_Overlap}.

\begin{figure*}[t]
\centering
{\subfloat[]{\includegraphics[width=2.7in]{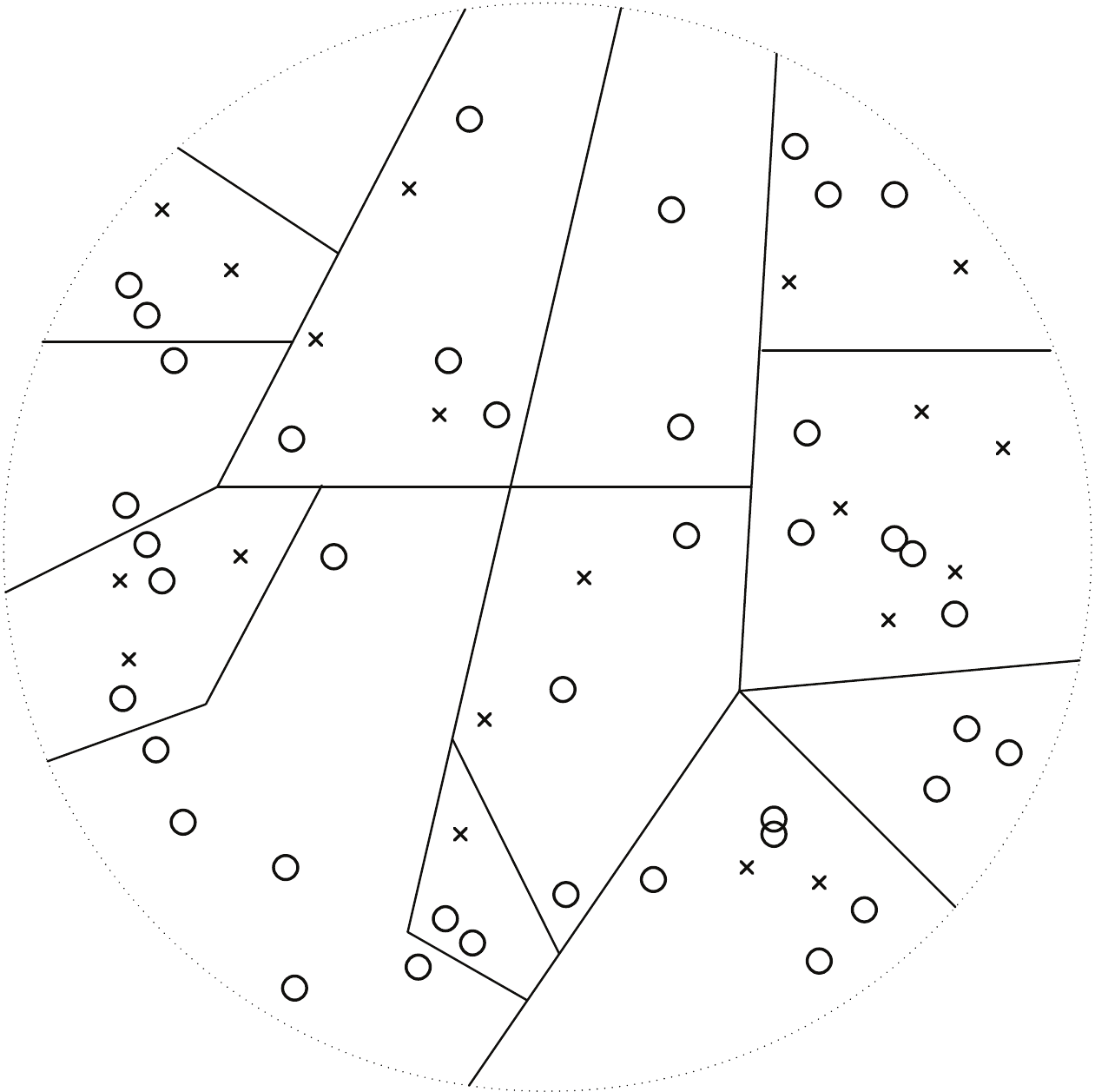}
\label{FIG_UGResults_1}}\hfil
\subfloat[]{\includegraphics[width=2.7in]{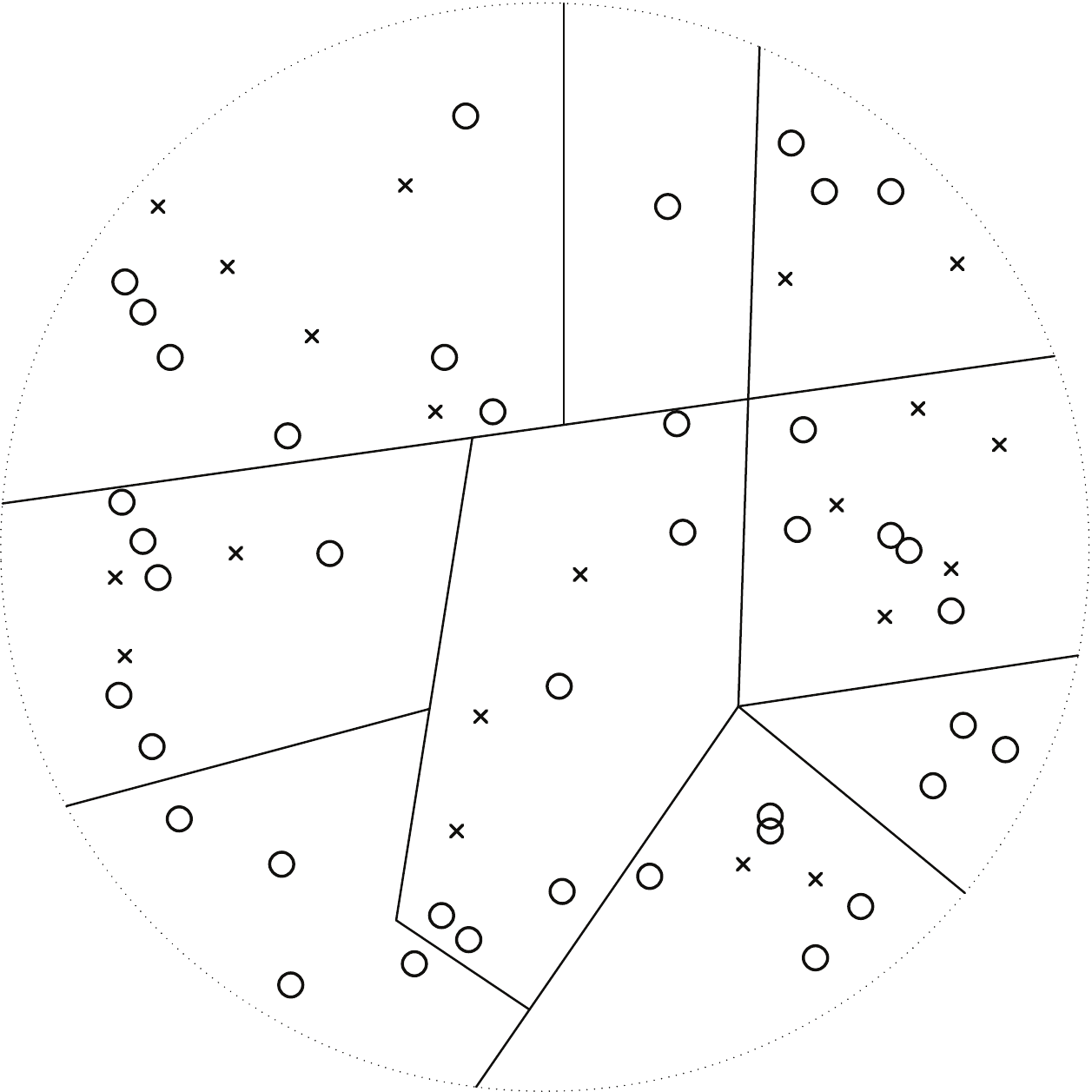}
\label{FIG_UGResults_2}}}
\caption{Illustration of user grouping results with different values of virtual cell size under the same topology as Fig. \ref{FIG_Snapshot_BS}. (a) $V=2$. (b) $V=3$. $\times$ represents a user and $\circ$ represents a distributed BS antenna. Different user groups are separated by solid lines. $L=40$. $K=20$.}
\label{FIG_UGResults}
\end{figure*}

Fig. \ref{FIG_UGResults} presents the grouping results by applying the proposed virtual-cell based user grouping algorithm to the topology given in Fig. \ref{FIG_Snapshot_BS}. It can be clearly seen that in contrast to Fig. \ref{FIG_Snapshot_BS} where the cluster-edge users suffer from strong interference from adjacent BSs, every user is always served by at least $V$ closest BS antennas. By doing so, all the users achieve relatively uniform rate performance with the lowest rate significantly improved compared to the BS clustering case, as we will demonstrate in the following subsection.

\subsection{Achievable Ergodic Rate $R_k$ with ZFBF}

Let us first extend the system model proposed in Section II to incorporate user grouping. Denote $\mathcal{K}_m$ as the user group that user $k$ belongs to, and $\mathcal{B}_m$ as the corresponding BS antenna set serving the users in $\mathcal{K}_m$, i.e., $\mathcal{B}_{m}=\cup_{j\in\mathcal{K}_{m}}\mathcal{V}_{j}$. The received signal of user $k$ can be then written as
\begin{equation}\label{signal model_G}
y_{k}=\mathbf{g}_{k,\mathcal{B}_{m}}\mathbf{x}_{k,\mathcal{B}_{m}}
+\sum_{j\in \mathcal{K}_{m}, j\neq k}\mathbf{g}_{k,\mathcal{B}_{m}}\mathbf{x}_{j, \mathcal{B}_{m}}
+\sum_{j\in \mathcal{K}_{t}, t\neq m}\mathbf{g}_{k,\mathcal{B}_{t}}\mathbf{x}_{j, \mathcal{B}_{t}}+n_{k}.
\end{equation}
In contrast to (\ref{signal model_NG}), the second item of (\ref{signal model_G}) denotes the intra-group interference, which can be effectively suppressed by coordinating transmissions of the virtual cells of users in the same group. In this section, we assume that ZFBF \cite{Caire} is adopted in each user group to eliminate the intra-group interference. For each user group $\mathcal{K}_{m}$, let $\mathbf{G}_{\mathcal{K}_{m}}\in\mathbb{C}^{|\mathcal{K}_{m}|\times|\mathcal{B}_{m}|}$ denote the channel gain matrix from user group $\mathcal{K}_{m}$ to its serving BS antenna set $\mathcal{B}_{m}$. $\mathbf{F}_{\mathcal{K}_{m}}$ denotes the pseudo-inverse of $\mathbf{G}_{\mathcal{K}_{m}}$, i.e, $\mathbf{F}_{\mathcal{K}_{m}}=\mathbf{G}_{\mathcal{K}_{m}}^{\dag}(\mathbf{G}_{\mathcal{K}_{m}}\mathbf{G}_{\mathcal{K}_{m}}^{\dag})^{-1}$. The precoding vector $\mathbf{w}_{k, \mathcal{B}_{m}}$ for any user $k\in\mathcal{K}_{m}$ can be then written as
\begin{align}\label{define of w_Z}
\mathbf{w}_{k,\mathcal{B}_{m}}=
\left\{ {\begin{array}{*{20}{c}}
{\frac{\mathbf{f}_{k, \mathcal{K}_{m}}}{\|\mathbf{f}_{k, \mathcal{K}_{m}}\|}}&{~~~\textmd{if}~~|\mathcal{B}_{m}|\geq |\mathcal{K}_{m}|}\\
{\mathbf{0}_{|\mathcal{B}_{m}|\times 1}}&{~~~\textmd{otherwise,} }
\end{array}} \right.
\end{align}
where $\mathbf{f}_{k, \mathcal{K}_{m}}$ is the column vector of $\mathbf{F}_{\mathcal{K}_{m}}$ corresponding to user $k$. Note that in (\ref{define of w_Z}), $\mathbf{w}_{k, \mathcal{B}_{m}}\neq \mathbf{0}$ only if $|\mathcal{B}_{m}|\geq |\mathcal{K}_{m}|$ because ZFBF requires that the number of transmit antennas is no smaller than the number of receive antennas. The achievable ergodic rate of user $k\in\mathcal{K}_m$ is then given by
\begin{equation}\label{Rk_Z}
R_{k}=\mathbb{E}_{\mathbf{H}_{\mathcal{K}_{m}}}
\left[\log_{2}\left(1+\frac{\bar{P}\mathbf{g}_{k,\mathcal{B}_{m}}\mathbf{w}_{k,\mathcal{B}_{m}}
\mathbf{w}_{k,\mathcal{B}_{m}}^{\dag}{\mathbf{g}}_{k,\mathcal{B}_{m}}^{\dag}}
{N_{0}+I_{k}}\right)\right],
\end{equation}
where $\mathbf{H}_{\mathcal{K}_{m}}\in\mathbb{C}^{|\mathcal{K}_{m}|\times|\mathcal{B}_{m}|}$ denotes the small-scale fading matrix from user group $\mathcal{K}_{m}$ to its serving BS antenna set $\mathcal{B}_{m}$, and the interference power $I_{k}$ is given by $I_{k}=\sum_{j\in\mathcal{K}_{t}, t\neq m}
\mathbb{E}\left[\mathbf{g}_{k,\mathcal{B}_{t}}\mathbf{w}_{j,\mathcal{B}_{t}}
\mathbf{w}_{j,\mathcal{B}_{t}}^{\dag}\mathbf{g}_{k,\mathcal{B}_{t}}^{\dag}\right] \bar{P}$.

\begin{figure}[t]
\begin{center}
\includegraphics[width=0.6\textwidth]{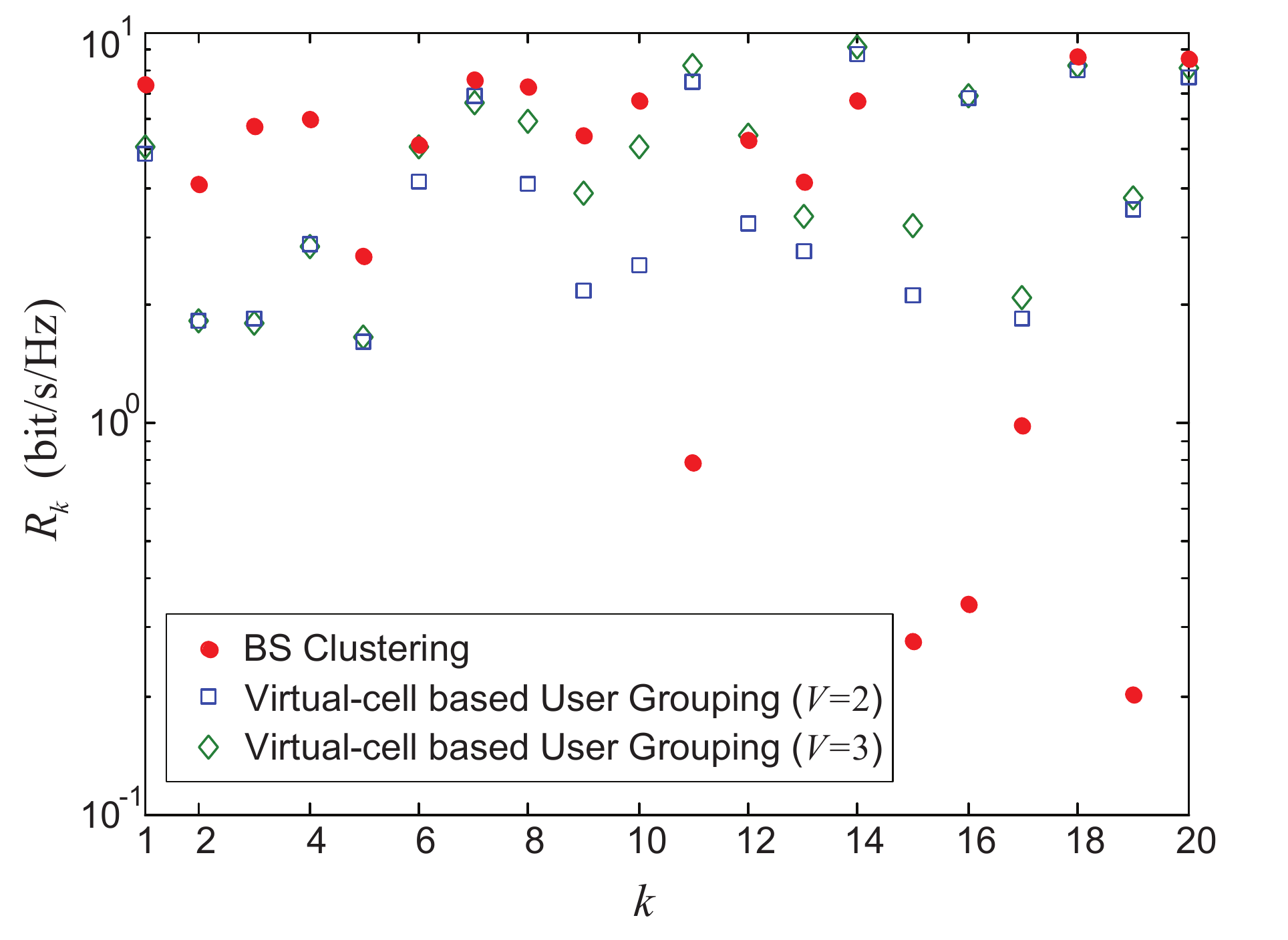}
\caption{Achievable ergodic rate of each user $R_{k}$ with virtual-cell based user grouping. The x-axis denotes the index of a user. $\alpha=4$. $\bar{P}/N_{0}=10$dB. $K=20$. $L=40$.}
\label{FIG_R_G}
\end{center}
\end{figure}

Fig. \ref{FIG_R_G} illustrates the achievable ergodic rate of each user under the topology given in Fig. \ref{FIG_UGResults}. For the sake of comparison, the achievable ergodic rate of each user with the BS clustering shown in Fig. \ref{FIG_Snapshot_BS} is also presented.\footnote{In this case, ZFBF is adopted in each BS cluster for joint transmission.} It can be clearly seen that the cluster-edge users in the BS clustering case achieve significantly lower rates than others. In contrast, with the virtual-cell based user grouping, the rate difference is greatly reduced, indicating that the rate performance of each user becomes much less sensitive to its position. It can be also observed from Fig. \ref{FIG_R_G} that by increasing the virtual cell size $V$, not only is the lowest rate improved, but also the user rate on average. In the next subsection, we will further study the effect of virtual cell size $V$ on the average user rate.

\begin{figure}[t]
\begin{center}
\includegraphics[width=0.6\textwidth]{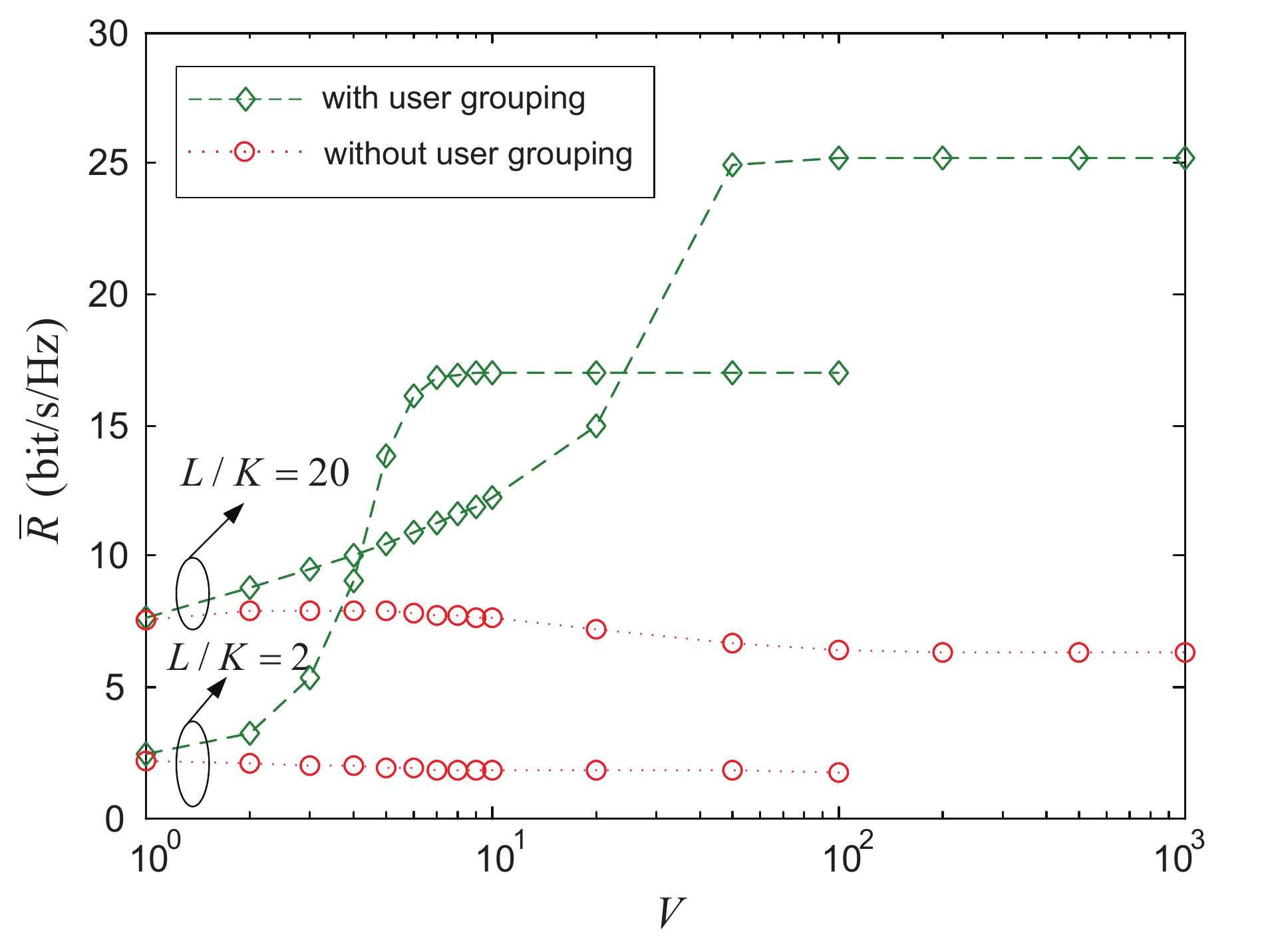}
\caption{Average user rate $\bar{R}$ with virtual-cell based user grouping versus virtual cell size $V$. $\alpha=4$. $\bar{P}/N_{0}=10$dB. $K=50$. $L/K=2, 20$.}
\label{FIG_OptV_G}
\end{center}
\end{figure}

\subsection{Effect of Virtual Cell Size $V$ on Average User Rate $\bar{R}$}

Similar to Section III, as the achievable ergodic rate of each user is closely dependent on their large-scale fading coefficients, we focus on the average user rate which is defined in (\ref{define of AverageR_NG}). Fig. \ref{FIG_OptV_G} shows the average user rate with the proposed virtual-cell based user grouping. For the sake of comparison, the rate performance without user grouping is also presented. It can be clearly seen from Fig. \ref{FIG_OptV_G} that in contrast to the no-grouping case where the highest average user rate is achieved when the virtual cell size $V$ is small, here the average user rate can be significantly improved with an increase in $V$.

Intuitively, by including more BS antennas into each user's virtual cell, more users would have overlapped virtual cells and be grouped together according to the virtual-cell based user grouping algorithm. With more users jointly served by using ZFBF to eliminate the intra-group interference, each of them achieves a higher rate thanks to a lower interference level. When the virtual cell size $V$ is large enough, nevertheless, all the users are grouped together, and the rate thus becomes saturated, as shown in Fig. \ref{FIG_OptV_G}. We can then conclude that here the virtual cell size $V$ determines a rate-complexity tradeoff: with a larger $V$, the average user rate is improved at the cost of a higher signal processing complexity as more users need to be jointly served.

\section{Conclusion}\label{sec:conclusion}

In this paper, the effect of virtual cell size on the average user rate of a large-scale downlink DAS is studied under different precoding schemes. The analysis shows that if MRT is adopted in each user's virtual cell, a small virtual cell size should be chosen so as to avoid sharing BS antennas for different users which would otherwise cause strong interference. On the other hand, if users are grouped with joint ZFBF transmission from their virtual cells to eliminate the intra-group interference, the average user rate could be significantly improved by increasing the virtual cell size. A novel virtual-cell based user grouping algorithm is proposed, with which the rate difference among users is greatly reduced compared to the conventional BS-centric clustering.

Note that despite the simplicity in concept, the group size, i.e., the number of users in each group, with the proposed virtual-cell based user grouping algorithm could be highly unbalanced if users' spatial distribution is nonuniform. For practical scenarios where users tend to congregate at some hot spots, it is important to further develop user grouping algorithms under certain complexity constraints such as the maximum group size. Moreover, in this paper, each user is assumed to select an identical number of closest BS antennas to form its virtual cell, which, though easy to implement in practice, may not be the optimal way for rate maximization. A global optimal antenna selection for all the users, on the other hand, would lead to prohibitively high computational complexity for a large-scale DAS. How to decompose the original combinatorial problem based on the concept of virtual cell is an interesting and challenging issue, which deserves much attention in the future study.

\appendices
\section{Derivation of (\ref{mu_NG})}\label{Ap_mu}
By substituting (\ref{define of I_NG}) into (\ref{define of mu_NG}), the average received SINR $\mu_{k}$ can be written as
\begin{align}\label{mu_ajl}
\mu_{k}=\frac{\|\bm{\gamma}_{k, \mathcal{V}_{k}}\|^2}
{\frac{N_{0}}{\bar{P}}+\sum_{j\in\mathcal{K}, j\neq k}\sum_{l\in\mathcal{V}_{j}}a_{j,l}\cdot\gamma_{k,l}^{2}}.
\end{align}
If the virtual cell size $V=1$, it is clear from (\ref{define of al_NG}) and (\ref{define of w_M}) that $a_{j,l}=1$. If $V>1$, by following a similar derivation to Appendix A in \cite{Junyuan}, $a_{j,l}$ can be obtained as
\begin{align}\label{ajl}
a_{j,l}=\sum_{m\in\mathcal{V}_{j}, m\neq l}
\frac{\beta_{j,l}^{-2}\beta_{j,m}^{-2}\left(\log\beta_{j,l}^{-2}-\log\beta_{j,m}^{-2}-1\right)+\beta_{j,m}^{-4}}
{\left(\beta_{j,l}^{-2}-\beta_{j,m}^{-2}\right)^2}\prod_{t\in\mathcal{V}_{j}, t\neq m, t\neq l}
\frac{\beta_{j,t}^{-2}}{\beta_{j,t}^{-2}-\beta_{j,m}^{-2}},
\end{align}
which can be further written as
\begin{align}\label{ajl1}
a_{j,l}=\sum_{m\in\mathcal{V}_{j}, m\neq l}
\frac{\gamma_{j,l}^{-2}\gamma_{j,m}^{-2}\left(\log\gamma_{j,l}^{-2}-\log\gamma_{j,m}^{-2}-1\right)+\gamma_{j,m}^{-4}}
{\left(\gamma_{j,l}^{-2}-\gamma_{j,m}^{-2}\right)^2}\prod_{t\in\mathcal{V}_{j}, t\neq m, t\neq l}
\frac{\gamma_{j,t}^{-2}}{\gamma_{j,t}^{-2}-\gamma_{j,m}^{-2}},
\end{align}
according to (\ref{define of beta_NG}).
(\ref{mu_NG}) can be then obtained by combining (\ref{mu_ajl}) and (\ref{ajl1}).

\section{Derivation of (\ref{AverageR_NG_ub})}\label{Ap_LogSI}

\begin{figure}[t]
\begin{center}
\includegraphics[width=0.4\textwidth]{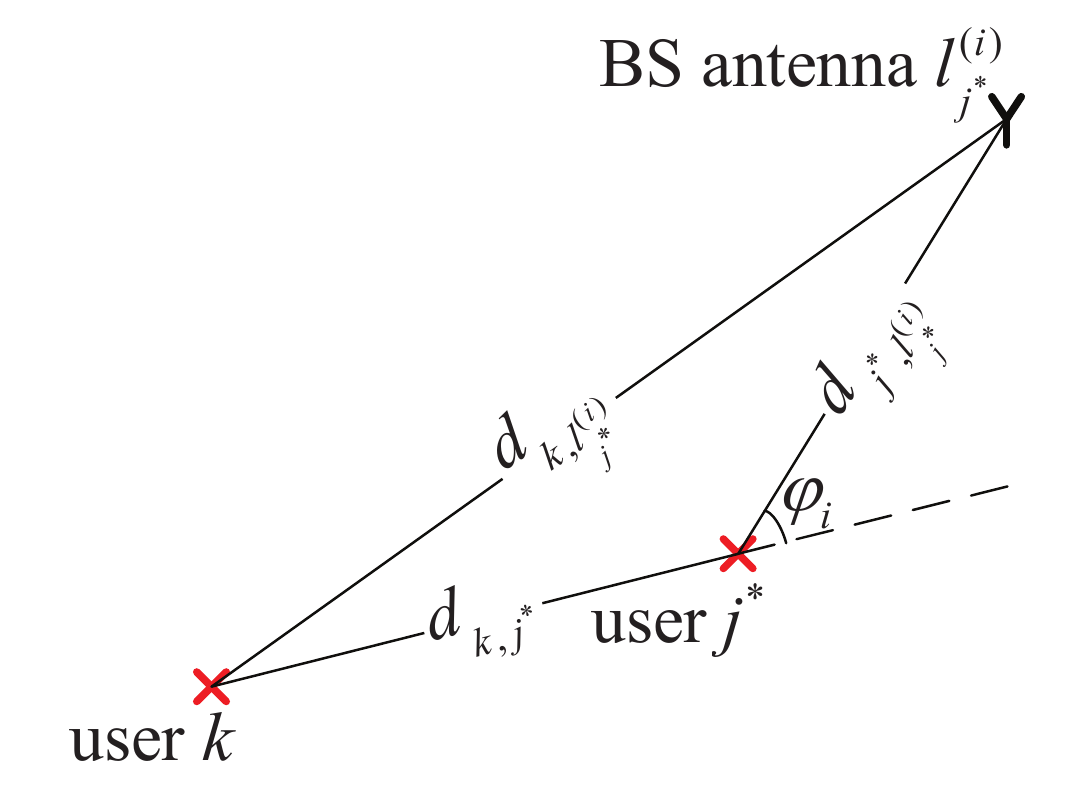}
\caption{Graphic illustration of $d_{k,l_{j^{*}}^{(i)}}$.}
\label{FIG_PDF_yi}
\end{center}
\end{figure}

According to (\ref{AverageR_mu_ub_NG_ub}), the upper-bound of the average user rate $\bar{R}^{ub}$ is determined by the normalized average received signal power $\tilde{S}_{k}=\|\bm{\gamma}_{k,\mathcal{V}_{k}}\|^2$ and the lower-bound of the normalized interference power $\tilde{I}_{k}^{lb}=\sum_{l\in\mathcal{V}_{j^{*}}}\Upsilon \left(\gamma_{j^{*}, l}; \left\{\gamma_{j^{*}, i}\right\}_{i\neq l, i\in\mathcal{V}_{j^{*}}}\right)\cdot \gamma_{k,l}^{2}$. Let
\begin{equation}\label{OderStatistics_L}
d_{k, l_{k}^{(1)}}\leq d_{k, l_{k}^{(2)}}\leq \cdots \leq d_{k, l_{k}^{(L)}}
\end{equation}
denote the order statistics obtained by arranging the access distances $d_{k, 1}, d_{k, 2}, \cdots ,d_{k, L}$ of user $k$ to $L$ BS antennas, where $l_{k}^{(i)}$ denotes the $i$th closest BS antenna of user $k$. Then the normalized average received signal power $\tilde{S}_{k}$ and the lower-bound of the normalized interference power $\tilde{I}_{k}^{lb}$ can be written as
\begin{equation}\label{S_ap}
\tilde{S}_{k}=\sum_{i=1}^{V}d_{k, l_{k}^{(i)}}^{-\alpha},
\end{equation}
and
\begin{align}\label{I_lb_ap}
\tilde{I}_{k}^{lb}=\sum_{i=1}^{V}\Upsilon\left(d_{j^{*}, l_{j*}^{(i)}}^{-\frac{\alpha}{2}}; d_{j^{*}, l_{j^{*}}^{(1)}}^{-\frac{\alpha}{2}},\cdots, d_{j^{*}, l_{j^{*}}^{(i-1)}}^{-\frac{\alpha}{2}}, d_{j^{*}, l_{j^{*}}^{(i+1)}}^{-\frac{\alpha}{2}}, \cdots, d_{j^{*}, l_{j^{*}}^{(V)}}^{-\frac{\alpha}{2}} \right)\cdot d_{k,l_{j^{*}}^{(i)}}^{-\alpha},
\end{align}
respectively, according to (\ref{define of gamma}), where $\Upsilon\left(x; b_{1}, b_{2}, \cdots, b_{V-1}\right)$ is given in (\ref{A}) and $d_{k, l_{j^{*}}^{(i)}}$ is the distance from user $k$ to the $i$th closest BS antenna of user $j^{*}$. Let $d_{k, j^{*}}$ denote the distance from user $k$ to user $j^{*}$. It can be easily obtained from Fig. \ref{FIG_PDF_yi} that
\begin{equation}\label{d_k_ljk_i}
d_{k, l_{j^{*}}^{(i)}}=\sqrt{d_{k, j^{*}}^2+d_{j^{*},l_{j^{*}}^{(i)}}^2+2d_{k, j^{*}}d_{j^{*},l_{j^{*}}^{(i)}}\cos{\varphi_{i}}}.
\end{equation}

\vspace{4mm}

Let $f_{d_{k, j^{*}}}(z)$ and $f_{\varphi_{i}}(\omega_{i})$ denote the pdfs of $d_{k, j^{*}}$ and $\varphi_{i}$, respectively, and $f_{d_{k,l_{k}^{(1)}}, d_{k, l_{k}^{(2)}}, \cdots, d_{k,l_{k}^{(V)}}}$\\$(x_{1}, x_{2}, \cdots, x_{V})$ denotes the joint pdf of $\left\{d_{k, l_{k}^{(i)}}\right\}_{i=1,\cdots,V}$. The upper-bound of the average user rate $\bar{R}^{ub}$ can be then written as
\begin{align}\label{AverageR_NG_ub_pdf}
\bar{R}^{ub}
=&\mathbb{E}_{\{\gamma_{j,l}\}_{j\in\mathcal{K},l\in\mathcal{B}}}\log_{2}\left[1+\frac{\tilde{S}_{k}}{\tilde{I}_{k}^{lb}}\right] \nonumber \\
=&\int_{0}^{1}f_{d_{k, j^{*}}}(z)
\underbrace{\int_{0}^{1}\int_{0}^{y_{V}}\cdots\int_{0}^{y_{2}}}_{V-fold}
f_{d_{j^{*},l_{j^{*}}^{(1)}}, d_{j^{*}, l_{j^{*}}^{(2)}}, \cdots, d_{j^{*},l_{j^{*}}^{(V)}}}(y_{1}, y_{2}, \cdots, y_{V}) \nonumber \\
&\underbrace{\int_{0}^{1}\int_{0}^{x_{V}}\cdots\int_{0}^{x_{2}}}_{V-fold}
f_{d_{k,l_{k}^{(1)}}, d_{k, l_{k}^{(2)}}, \cdots, d_{k,l_{k}^{(V)}}}(x_{1}, x_{2}, \cdots, x_{V}) \underbrace{\int_{0}^{2\pi}\int_{0}^{2\pi}\cdots\int_{0}^{2\pi}}_{V-fold}\prod_{i=1}^{V}f_{\varphi_{i}}(\omega_{i})
\nonumber \\
&\log_{2}\left(1+\frac{\sum_{i=1}^{V}x_{i}^{-\alpha}}{\sum_{i=1}^{V}\Upsilon\left(y_{i}^{-\frac{\alpha}{2}}; y_{1}^{-\frac{\alpha}{2}},\cdots, y_{i-1}^{-\frac{\alpha}{2}}, y_{i+1}^{-\frac{\alpha}{2}}, \cdots, y_{V}^{-\frac{\alpha}{2}} \right)\cdot \left(z^{2}+y_{i}^{2}+2y_{i}z\cos\omega_{i}\right)^{-\frac{\alpha}{2}}}\right) \nonumber \\ &d\omega_{1}d\omega_{2}\cdots d\omega_{V}dx_{1}dx_{2}\cdots dx_{V}dy_{1}dy_{2}\cdots dy_{V} dz,
\end{align}
by combining (\ref{AverageR_mu_ub_NG_ub}) and (\ref{S_ap}-\ref{d_k_ljk_i}).
Similarly, we have
\begin{align}\label{E_log_S_pdf}
\mathbb{E}_{\{\gamma_{k,l}\}_{k\in\mathcal{K}, l\in\mathcal{B}}}\log_{2}\tilde{S}_{k}
{=}&\underbrace{\int_{0}^{1}\hspace{-1mm}\int_{0}^{x_{V}}\hspace{-1mm}\int_{0}^{x_{V-1}}\hspace{-1mm}\cdots\int_{0}^{x_{2}}}_{V-fold}
\hspace{-1mm}\log_{2}\left(\sum_{i=1}^{V}x_{i}^{-\alpha}\right)f_{d_{k,l_{k}^{(1)}}, d_{k, l_{k}^{(2)}}, \cdots, d_{k,l_{k}^{(V)}}}(x_{1}, x_{2}, \cdots, x_{V}) \nonumber \\
&dx_{1}dx_{2}\cdots dx_{V},
\end{align}
and
\begin{align}\label{E_logI_lb_pdf}
\mathbb{E}_{\{\gamma_{k,l}\}_{k\in\mathcal{K}, l\in\mathcal{B}}}\log_{2}\tilde{I}_{k}^{lb}
&=\int_{0}^{1}f_{d_{k, j^{*}}}(y)
\underbrace{\int_{0}^{1}\int_{0}^{x_{V}}\cdots\int_{0}^{x_{2}}}_{V-fold}
f_{d_{j^{*},l_{j^{*}}^{(1)}}, d_{j^{*}, l_{j^{*}}^{(2)}}, \cdots, d_{j^{*},l_{j^{*}}^{(V)}}}(x_{1}, x_{2}, \cdots, x_{V}) \nonumber \\
&\underbrace{\int_{0}^{2\pi}\hspace{-1mm}\int_{0}^{2\pi}\hspace{-1mm}\cdots\hspace{-1mm}\int_{0}^{2\pi}}_{V-fold} \prod_{i=1}^{V}f_{\varphi_{i}}(\omega_{i})
\log_{2}\hspace{-1mm}\left(\sum_{i=1}^{V}\Upsilon\left(x_{i}^{-\frac{\alpha}{2}}; x_{1}^{-\frac{\alpha}{2}},\cdots, x_{i-1}^{-\frac{\alpha}{2}}, x_{i+1}^{-\frac{\alpha}{2}}, \cdots, x_{V}^{-\frac{\alpha}{2}} \right)\right. \nonumber \\
&\left.\cdot \left(y^{2}+x_{i}^{2}+2x_{i}y\cos\omega_{i}\right)^{-\frac{\alpha}{2}}\right)
d\omega_{1}d\omega_{2}\cdots d\omega_{V}dx_{1}dx_{2}\cdots dx_{V}dy.
\end{align}

Let us ignore the edge effect and assume that an infinite number of BS antennas and users are uniformly distributed in an infinitely large plane with $L$ BS antennas and $K$ users in a disk with radius 1. It can be easily obtained that the pdf and the cumulated density function (cdf) of the access distance from user $k$ to BS antenna $l$ are given by
$f_{d_{k,l}}(x)=2x$ and $F_{d_{k,l}}(x)=x^{2}$, respectively. We then have
\begin{align}\label{JointPdf_d_{k}}
f_{d_{k,l_{k}^{(1)}}, d_{k, l_{k}^{(2)}}, \cdots, d_{k,l_{k}^{(V)}}}(x_{1}, x_{2}, \cdots, x_{V})
=\frac{L!}{(L-V)!}\left(1-x_{V}^{2}\right)^{L-V}\prod_{i=1}^{V}2x_{i},
\quad x_{1}\leq x_{2}\leq\cdots \leq x_{V},
\end{align}
\begin{equation}\label{pdf of djmin}
f_{d_{k, j^{*}}}(x)=(K-1)(1-x^{2})^{K-2}\cdot 2x.
\end{equation}
and
\begin{align}\label{pdf of phi_ap}
f_{\varphi_{i}}(x_{i})=\frac{1}{2\pi}.
\end{align}
(\ref{AverageR_NG_ub}) can be then obtained by substituting (\ref{JointPdf_d_{k}}-\ref{pdf of phi_ap}) into (\ref{AverageR_NG_ub_pdf}).

Moreover, we have
\begin{align}\label{E_log_S}
\mathbb{E}_{\{\gamma_{k,l}\}_{k\in\mathcal{K}, l\in\mathcal{B}}}\log_{2}\tilde{S}_{k}
=&\frac{2^{V}L!}{(L-V)!}\underbrace{\int_{0}^{1}\int_{0}^{x_{V}}\int_{0}^{x_{V-1}}\cdots\int_{0}^{x_{2}}}_{V-fold}
\log_{2}\left(\sum_{i=1}^{V}x_{i}^{-\alpha}\right)\left(1-x_{V}^{2}\right)^{L-V} \nonumber \\
&\prod_{i=1}^{V}{x}_{i}dx_{1}dx_{2}\cdots dx_{V},
\end{align}
and $\mathbb{E}_{\{\gamma_{k,l}\}_{k\in\mathcal{K}, l\in\mathcal{B}}}\log_{2}\tilde{I}_{k}^{lb}$ can be written as
\begin{align}\label{E_logI_lb}
\mathbb{E}_{\{\gamma_{k,l}\}_{k\in\mathcal{K}, l\in\mathcal{B}}}\log_{2}\tilde{I}_{k}^{lb}
=&\frac{2(K-1)L!}{\pi^{V}(L-V)!}\int_{0}^{1}y\left(1-y^{2}\right)^{K-2}
\underbrace{\int_{0}^{1}\int_{0}^{x_{V}}\int_{0}^{x_{V-1}}\cdots\int_{0}^{x_{2}}}_{V-fold}
\left(1-x_{V}^{2}\right)^{L-V}\prod_{i=1}^{V}x_{i} \nonumber \\
&\underbrace{\int_{0}^{2\pi}\int_{0}^{2\pi}\cdots\int_{0}^{2\pi}}_{V-fold}
\log_{2}\left(\sum_{i=1}^{V}\Upsilon\left(x_{i}^{-\frac{\alpha}{2}}; x_{1}^{-\frac{\alpha}{2}}, \cdots, x_{i-1}^{-\frac{\alpha}{2}}, x_{i+1}^{-\frac{\alpha}{2}}, \cdots, x_{V}^{-\frac{\alpha}{2}} \right)\cdot \right. \nonumber \\
&\left.\left(x_{i}^{2}+y^{2}+2x_{i}y\cos\omega_{i}\right)^{-\frac{\alpha}{2}}\right)
d\omega_{1}d\omega_{2}\cdots d\omega_{V}dx_{1}dx_{2}\cdots dx_{V}dy,
\end{align}
by substituting (\ref{JointPdf_d_{k}}-\ref{pdf of phi_ap}) into (\ref{E_log_S_pdf}) and (\ref{E_logI_lb_pdf}), respectively.

\section{Derivation of (\ref{OptV_NG})}\label{Ap_OptV}
Let us rewrite the optimization problem formulated in (\ref{Opt_Prob}) and (\ref{constraint_relax}) as
\begin{align}\label{Ap1}
&\text{max} \qquad V \\ 
&\text{s.t.} \; \qquad \bar{r}_{k}+\bar{r}_{j^{*}}\leq \bar{d}_{k,j^{*}}, \label{Ap2}
\end{align}
where $\bar{r}_{k}=\mathbb{E}_{\{\gamma_{j,l}\}_{j\in\mathcal{K}, l\in\mathcal{B}}}\left[r_{k}\right]$ and $\bar{r}_{j^{*}}=\mathbb{E}_{\{\gamma_{j,l}\}_{j\in\mathcal{K}, l\in\mathcal{B}}}\left[r_{j^{*}}\right]$.
As $L$ BS antennas are uniformly distributed, we have
\begin{align}\label{Ap3}
\bar{r}_{k}^{2}=\bar{r}_{j^{*}}^{2}=\frac{V}{L}.
\end{align}
The constraint (\ref{Ap2}) can be then written as
\begin{equation}\label{Ap4}
V\leq \frac{L}{4}\left(\bar{d}_{k, j^{*}}\right)^{2}.
\end{equation}
It is clear from (\ref{Ap1}) and (\ref{Ap4}) that the optimal virtual cell size $V^{*}$ is
\begin{align}\label{Ap5}
V^{*}=\frac{L}{4}\left(\bar{d}_{k, j^{*}}\right)^{2}.
\end{align}
Note that the pdf of the distance from user $k$ to its closest user $j^{*}$, ${d}_{k, j^{*}}$, has been given in (\ref{pdf of djmin}). We then have
\begin{align}\label{Avereagedjk*_ap_int}
\bar{d}_{k,j^{*}}&=2\int_{0}^{1}x^{2}(K-1)\left(1-x^{2}\right)^{K-2}dx \nonumber \\
&\mathop{\approx}^{\text{for large }K}2\int_{0}^{1}x^{2}(K-1)e^{-(K-2)x^{2}}dx \nonumber \\
&\mathop{=}^{y=(K-2)x^2}
\frac{K-1}{(K-2)^{\frac{3}{2}}}\int_{0}^{K-2}y^{\frac{1}{2}}e^{-y}dy \nonumber \\
&=\frac{K-1}{(K-2)^{\frac{3}{2}}}\left\{-(K-2)^{\frac{1}{2}}e^{-(K-2)}+\frac{1}{2}\Gamma\left(\frac{1}{2}, 0\right)-\frac{1}{2}\Gamma\left(\frac{1}{2}, K-2\right)\right\},
\end{align}
where $\Gamma(s,x)=\int_{x}^{\infty}t^{s-1}e^{-t}dt$. Note that for large $K$, $(K-2)^{\frac{1}{2}}e^{-(K-2)}\approx 0$, $\Gamma\left(\frac{1}{2}, K-2\right)\approx 0$ and $\frac{(K-1)}{(K-2)^{\frac{3}{2}}}\approx \frac{1}{\sqrt{K}}$. (\ref{Avereagedjk*_ap_int}) can be then approximated as
\begin{align}\label{Avereagedjk*_ap_Gamma}
\bar{d}_{k,j^{*}}\approx\frac{1}{2}\Gamma\left(\frac{1}{2}, 0\right)\cdot\frac{1}{\sqrt{K}}
\approx 0.89\frac{1}{\sqrt{K}}.
\end{align}
Finally, (\ref{OptV_NG}) can be obtained by combining (\ref{Ap5}-\ref{Avereagedjk*_ap_Gamma}).

\end{document}